\documentclass[twocolumn]{revtex4-2}
\pdfoutput=1
\usepackage{graphicx} 
\usepackage[english]{babel}
\usepackage{titlesec}
\titlespacing*{\section}
{0pt}{2ex plus 1ex minus .2ex}{0ex plus .2ex}
\titlespacing*{\subsection}
{0pt}{1ex plus 1ex minus .2ex}{0ex plus .2ex}

\usepackage{natbib}
\setcitestyle{super}
\usepackage{times}
\usepackage{amsmath} 
 
\usepackage{rotating}

\makeatletter
\renewcommand*{\fnum@figure}{{\normalfont\bfseries \figurename~\thefigure}}
\renewcommand*{\@caption@fignum@sep}{\textbf{ : }}
\makeatother

\makeatletter
\renewcommand*{\fnum@table}{{\normalfont\bfseries \tablename~\thetable}}
\makeatother

\begin{document}

\title{Novel Liquid-Liquid Interface Deposition Method for Thin Films of Two-Dimensional Solids}
\author{Amy R. Smith$^{1*}$}
\author{Muhammad Zulqurnain$^{1,2,3}$}
\author{Angus G.M. Mathieson$^{1,3}$}
\author{Marek Szablewski$^{1*}$}
\author{Michael R.C. Hunt$^{1*}$}

\affiliation{$^1$Department of Physics, Durham University, South Road, Durham, DH1 3LE, U.K.}
\affiliation{$^2$Current address: Cavendish Laboratory, Department of Physics, University of Cambridge, Cambridge, CB3 0HE, U.K.}
\affiliation{$^3$These authors contributed equally: Muhammad Zulqurnain, Angus G.M. Mathieson.}
\affiliation{$^*$amy.r.smith@durham.ac.uk, marek.szablewski@durham.ac.uk, m.r.c.hunt@durham.ac.uk}

\begin{abstract}
    Thin films and van der Waals heterostructures (vdWHs) derived from two-dimensional solids offer enormous potential for a broad range of novel, energy efficient devices, however their use is currently hampered by slow, labor-intensive fabrication methods often employing hazardous chemicals. We demonstrate a novel liquid-liquid interface technique for rapid, low-cost, and environmentally-friendly production of ultra-thin films and vdWHs of two-dimensional solids from aqueous surfactant-stabilized suspensions. The approach is generic to two-dimensional materials which can be stabilized in aqueous suspension by a surfactant and the resulting films can be transferred to an arbitrary substrate by a range of approaches. The wide applicability of this technique is demonstrated through production of thin films on a variety of substrates, deposition of transparent, highly conductive graphene films with conductivities between $7.7 \times 10^{3} - 1.26\times 10^{5}~\textrm{S~m}^{-1}$ and transmittances of 55-75\%, and by the fabrication of a vdWH of MoS$_2$, WS$_2$, and few-layer graphene.
\end{abstract}

\maketitle

Thin films of two-dimensional (2D) solids and van der Waals heterostructures (vdWHs) derived from them offer enormous potential for a broad range of applications, including transistors\cite{Schwierz2010,Schwierz2011,Radisavljevic2011,Hopf2014, Rhee2024}, solar cells\cite{Wang2008,ricciardulli2020}, and functional devices and coatings\cite{ Varrla2015, Li2018, De2010,Matkovic2016, Woltornist2017,carey2024}. vdWHs in particular leverage the lack of dangling bonds at the surface of a 2D solid to produce atomically-abrupt and defect-free interfaces without the constraints of lattice-matching demanded by conventional heterostructures\cite{Geim2013}. Novel structures based on 2D building blocks can, in principle, be assembled arbitrarily, however, in practice, methods of producing thin films and vdWHs of 2D materials are seriously constrained by laborious, non-scalable `physical' methods such as micromechanical exfoliation\cite{Castellanos-Gomez2014} or non-generic approaches\cite{kim2022,zhang2020} which are often slow and/or involve the extensive use of environmentally-unfriendly chemicals. 

Here we report a novel liquid-liquid interface\cite{Woltornist2013, Clark2017, Yu2015, Neilson2020, cassidy2025} technique for the controllable formation and deposition of monolayer and few-layer films of 2D solids and van der Waals heterostructures derived from them. The approach, which we term `Liquid Interface Deposition' (LID), is simple, rapid (approximately 7 hours from starting solid bulk `parent' materials to deposited film), scalable, removes surfactants/thicker material, minimizes the use of hazardous chemicals, and recycles process chemicals. LID enables deposition of films of 2D materials on substrates without prior surface treatment, employing a surfactant-stabilized aqueous suspension of thin platelets of the 2D solid as a precursor. Moreover, by successive deposition vdWHs may be rapidly produced, which, to the best of our knowledge, is the first demonstration of the deposition of vdWHs, as opposed to heterogeneous single layer films\cite{Clark2017}, by liquid-liquid interfacial deposition. The technique reported here is applicable to any 2D solid which can be dispersed in surfactant-stabilized aqueous suspension and therefore includes the vast majority of those materials which do not require a substrate or other external `scaffold' or support for stability (e.g., two-dimensional layered materials (2DLMs) such as graphene, transition metal dichalcogenides (TMDCs) etc.\cite{Geim2013, Rhee2024}).

The LID method performs three roles: first, the assembly of a thickness-controlled film of platelets of a 2D solid at the interface of two immiscible liquids, which can then be readily transferred onto a substrate through dipping, as discussed here. (Other transfer approaches such as horizontal dipping, pouring, or freeze transfer are discussed in the Supplementary Information, Section S1); second, the removal of platelets of undesirable (i.e., multilayer) thickness from the film, if present in the initial surfactant-stabilized suspension; finally and crucially, the removal of surfactant from the platelets, forming a `clean', uncontaminated film of the 2D solid(s). Vertical vdWHs are formed by repeating the process multiple times using different 2D solids without measurable disruption to previously deposited layers.The LID approach reported here, has significant advantages over previously reported liquid-liquid interfacial assembly and deposition mechanisms,\cite{Woltornist2013, Yu2015, Clark2017, Neilson2020} with its rapidity, the capacity to remove surfactants/contaminants, and broad applicability (see Supplementary Information Table S1 for comparison with other methods). Moreover, both the production of precursor suspension and the deposition method are readily scaled to mass production,\cite{paton2014,bigcentrifuges} have a low energy budget, require little specialized equipment or technical know-how, can be adapted to ensure minimum waste and environmental impact, and no direct chemical modification is made to the 2D solids in any part of the process. 

\section*{\raggedright{Results and Discussion}}
\subsection*{\raggedright{Liquid Interface Deposition}}
Surfactant-stabilized aqueous suspensions of thin (predominantly monolayer to few layer) platelets of a chosen 2D material were used as precursors for film and vdWH deposition. The suspensions were produced by a common, rapid approach for all the 2DLMs studied, as described in the methods section, obviating the need for specific liquid phases\cite{shen2015liquid} and protocols\cite{kim2022} for each material. Triton X-100 was chosen as a low-cost, readily available surfactant, although the approach outlined can be readily modified to accommodate a broad range of common surfactants (see Supplementary Information, Figure S4). 

\begin{figure}[ht!]
    \centering
    \includegraphics[width=\linewidth]{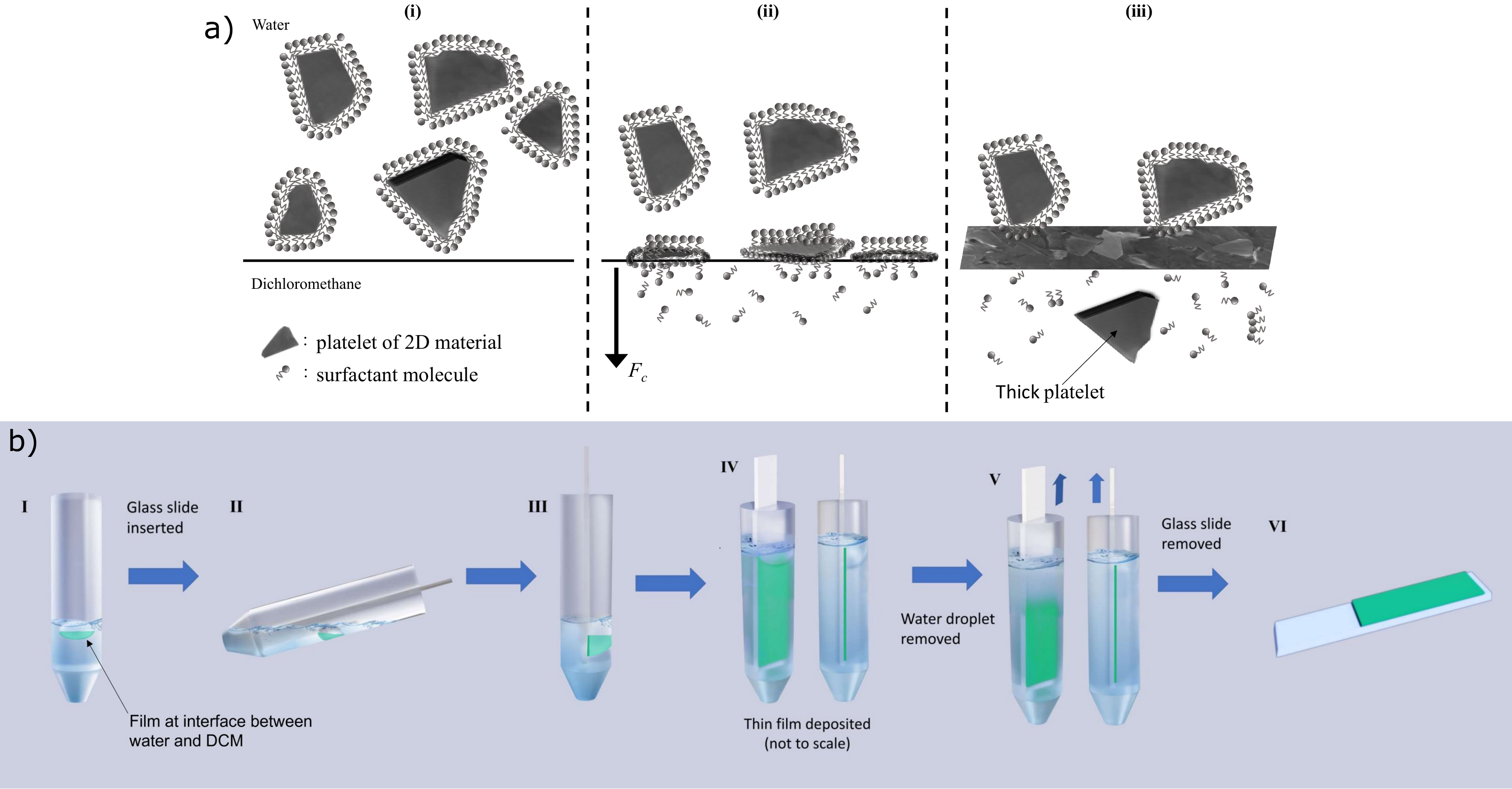}
    \caption{\textit{a) Proposed mechanism for formation of a 2DLM platelet film at the interface between the two immiscible liquids (water and dichloromethane for the data presented in this work) with the surfactant being removed from the 2DLM platelets into the denser liquid phase (dichloromethane). (i) State of system before centrifuging. (ii) Surfactant-stabilized 2DLM platelets  migrate to the interface under the action of centripetal forces generated during centrifuging. (iii) Surfactant molecules and thicker platelets preferentially move across the interface into the denser solvent, while the 2DLM platelets assemble at the interface to minimize surface energy and form a film. b) Illustration of the process used for the deposition of thin films and heterostructures. A centrifuge tube containing a 2DLM platelet film assembled at the interface between two immiscible liquids (I) (water and dicloromethane) was tilted, and the substrate inserted (II). The centrifuge tube was then returned to the upright (III). Further separation solvent was added to the tube, raising the interface and depositing the film onto the substrate (IV), after which the water droplet was removed (V) followed by the substrate (VI).}}
    \label{fig:1}
\end{figure}

To facilitate the LID process, the precursor suspension of 2DLM platelets was added to a centrifuge tube containing a `separation solvent'. The separation solvent was chosen such that: (1) it is immiscible with and denser than the aqueous suspension; (2) it has a high partition coefficient such that the surfactant prefers to be in the separation solvent to the aqueous phase; (3) the surfactant can be readily removed enabling the separation solvent to be re-cycled and re-used. In the experiments reported in this work, dichloromethane (DCM) was selected as meeting these conditions, with the partition coefficient, $K_{D/W}$, of Triton X-100 between DCM and water measured to be $220\pm70$ (in preparation, Finlay R. Dover, A.R.S., M.S., M.R.C.H.). In principle, any solvent/surfactant which meets the criteria described should be suitable for LID. This mixture is then centrifuged or left to settle via gravity (Supplementary Information, Figure S5), after which a thin film of 2DLM platelets is formed at the interface of the DCM and aqueous phase. Our proposed mechanism for the assembly of 2DLM platelets at the liquid-liquid interface is shown schematically in Figure \ref{fig:1}a).

Film deposition is shown schematically in Figure \ref{fig:1}b). To ensure reproducibility and consistency between films this process was automated by using a custom-built programmable dipping system (Supplementary Information, Figure S6 and video) such that insertion angle, dipping speed etc. could be fully controlled. However, it is also possible to carry out the process by hand. The dipping method may be repeated multiple times with the same material to build up thick films and with different materials to produce vdWHs, with no measurable damage to, or visible loss of, previously deposited material on subsequent dips. 

\subsection*{\raggedright{Versatility}}
\begin{figure}[ht!]
    \centering
    \includegraphics[width=\linewidth]{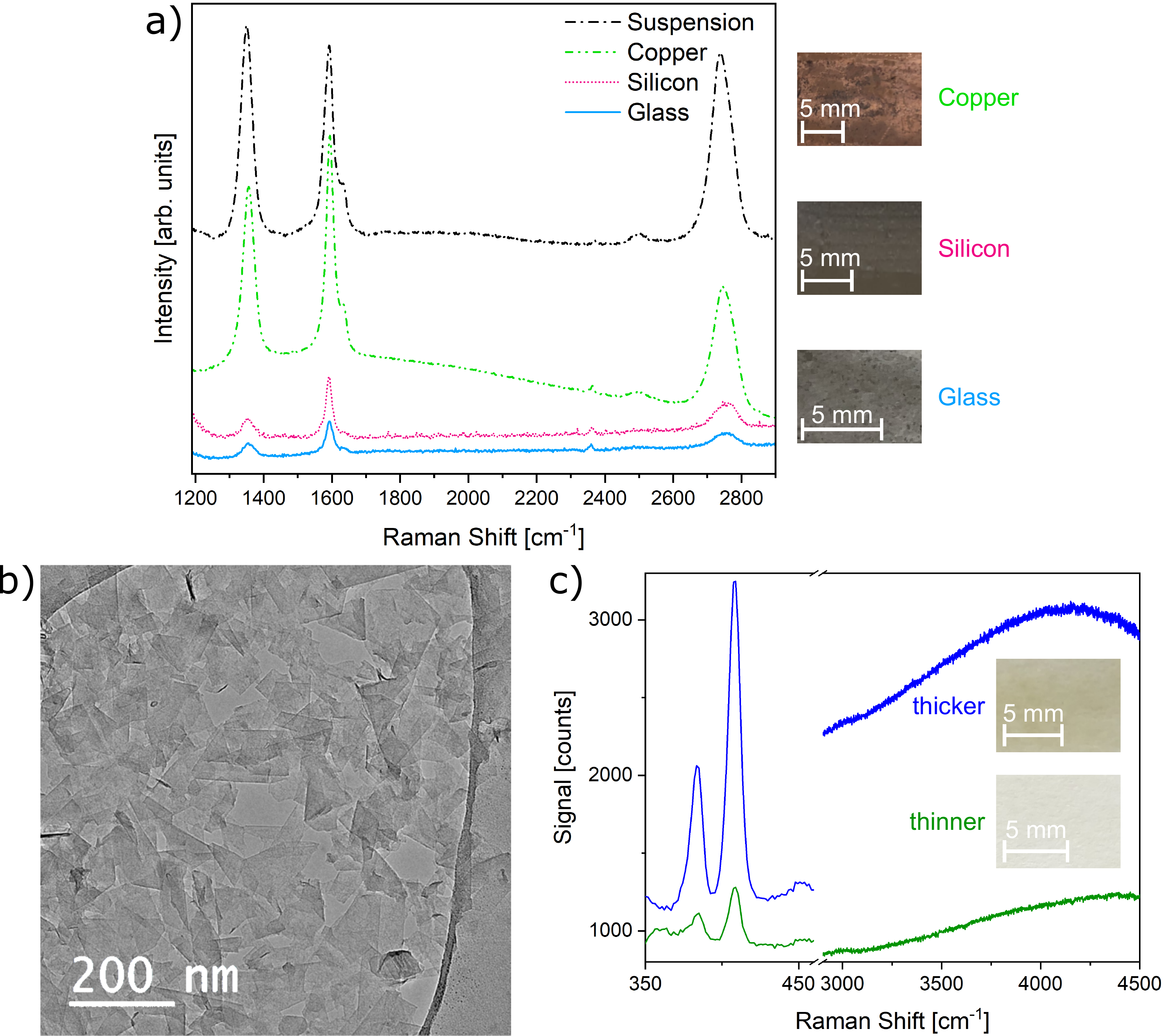}
    \caption{\textit{a) Raman spectra of few-layer graphene films deposited on different substrates and the surfactant-stabilized few-layer graphene aqueous suspension used as a precursor for deposition. Similar lineshapes between the few-layer graphene suspension and the films deposited on all substrates indicate no detectable increase in defect density upon formation of the film and film deposition. The data have been offset vertically to allow for easier comparison. b) Transmission electron microscopy (TEM) image showing a few-layer graphene film deposited onto a TEM grid via the LID method. The film can be seen to be largely continuous with only a few pinholes. c) Raman spectra of two films of MoS$_2$ of different thicknesses, with the thinner sample shown in green and the thicker shown in blue. The E$^{1}_{2g}$ and A$_{1g}$ peaks characteristic of MoS$_2$ can be clearly seen in both spectra, as can the $A$ exciton peak. The position of the $A$ exciton peak shows that these films are not behaving as in the bulk regime.}}
    \label{fig:2}
\end{figure}

The applicability of the LID process to a broad range of substrates is demonstrated in Figure \ref{fig:2}a), which presents Raman spectra and optical images of few-layer graphene (FLG) films deposited on copper, silicon, and (hydrophilic) glass. The films displayed a high degree of uniformity both optically and between regions selected for acquisition of the Raman spectra. Comparison between the Raman spectra of the thin films and that of the parent surfactant-stabilized aqueous suspension, also shown in Figure \ref{fig:2}a), indicates similar characteristic Raman peaks, showing that no additional defects are introduced upon deposition. The continuity of the FLG films is demonstrated in Figure \ref{fig:2}b), which shows a transmission electron microscopy (TEM) image of a graphene film deposited on a holey carbon grid (a TEM image for a MoS$_2$ film is shown in Supplementary Information, Figure S7). Although the film was produced using a single dipping cycle, and is therefore thin, it can be seen to be largely continuous, consisting of overlapping FLG platelets with very few pinholes present. We find the continuous regions of the films produced by the LID process are substantially larger than those produced by Langmuir-Blodgett deposition by ourselves (Supplementary Information, Figures S8 and S9), and similar to those produced by other liquid-liquid interface deposition methods\cite{cassidy2025}.

The thickness of the films produced can be altered by altering the concentration of the surfactant-stabilized aqueous suspension of 2DLM, by repeating the dipping process multiple times, or both. The minimum film thickness is, in principle, limited by the minimum thickness of platelets in suspension. The maximum film thickness which may be achieved in a single dip is limited by the suspension concentration. These factors are dependent on the 2DLM and the precise preparation method (e.g. any size/thickness selection of the platelets in the suspension prior to deposition), however the LID method remains consistent for all 2DLMs and preparation methods. Raman spectra of films produced using the multiple dipping approach are shown in Figure \ref{fig:2}c) for two MoS$_2$ samples produced using different numbers of dips. No change in peak separation, or relative intensity of the E$^{1}_{\text{2g}}$ and A$_{\text{1g}}$ modes is observed (Figure \ref{fig:2}c)). The separation of the E$^{1}_{\text{2g}}$ and A$_{\text{1g}}$ is found to be 24~cm$^{-1}$ in both films, corresponding to platelets of 4 to 5 layers\cite{Lee2010}, indicating a weak interaction between the platelets as the film thickness is increased. A peak corresponding to the $A$ exciton\cite{Mak2010} can be seen at 1.8~eV, which increases in intensity in the thicker film, indicating that a proportion of the platelets consist of a smaller number of layers and supporting the suggestion of weak vertical coupling between them.

\begin{figure}[ht]
    \centering
    \includegraphics[width=\linewidth]{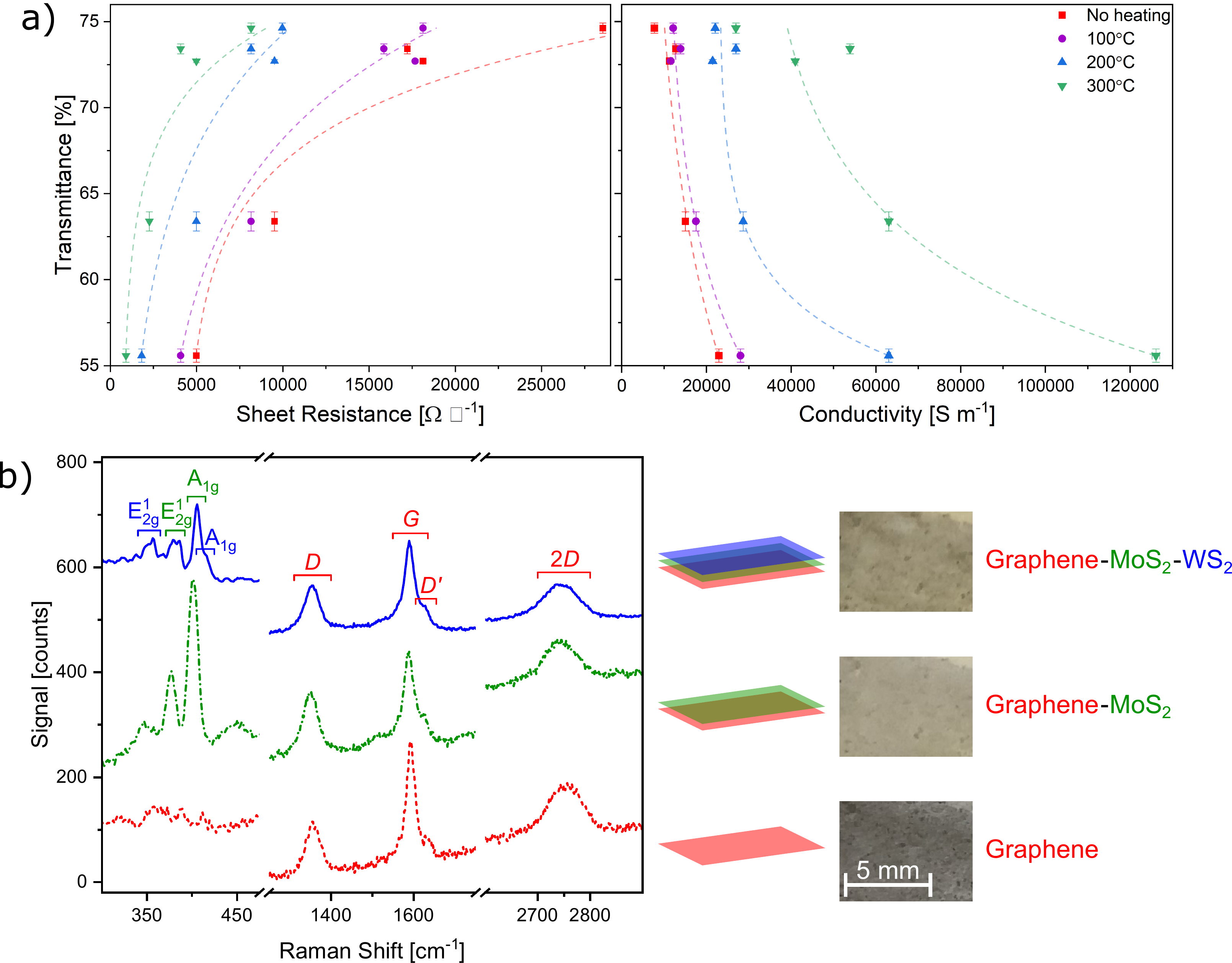}
    \caption{ \textit{a) Sheet resistance and conductivity of few-layer graphene (FLG) films of different optical transmittance both as-deposited and after annealing for 24 hours at different temperatures. The dashed lines are guides to the eye. Error bars are standard errors. b) Raman spectra of a FLG-MoS$_2$-WS$_2$ van der Waals heterostructure (vdWH) deposited on glass at each stage of fabrication. The red (dashed) data correspond to the first (FLG) layer, the green (dash-dot) data to a FLG-MoS$_2$ vdWH, and the blue data to a FLG-MoS$_2$-WS$_2$ vdWH. The data have been offset vertically to allow for easier comparison. Features from MoS$_2$, WS$_2$ and graphene are present in the Raman spectra of the final vdWH, indicating the vdWH fabrication has been successful. The photographs of the vdWH in each stage of fabrication are from the same area of the sample}.}
    \label{fig:3}
\end{figure}

\subsection*{\raggedright{Transparent conductive few-layer graphene films}}
FLG films were deposited onto glass slides to form continuous transparent films with absorption which was found to vary weakly over visible optical wavelengths\cite{Nair2008}. Sheet resistance was measured and found to decrease with transmittance, Figure \ref{fig:3}a), as would be expected from the accompanying increase in film thickness\cite{liu2013electrochemically}. Films were annealed under an inert (argon) atmosphere at temperatures between 100 and 300~{\textdegree}C which resulted in an enhancement of conductivity, increasing with anneal temperature. This behavior may be attributed to the removal of residual water and other contaminants which impede charge transfer between the FLG platelets and/or structural re-arrangement within the film leading to a greater contact between the platelets. Annealing films produced by drop-casting surfactant-stablized suspensions is found to lead to decomposition and not removal of surfactant hence the presence and subsequent removal of residual Triton X-100 cannot explain the observed improvement of charge transport of the LID FLG films, and is consistent with surfactant removal during film formation. The relationship between transmittance and sheet resistance for these FLG films does not follow the typical relationship for thin metallic films, but instead shows percolative-like behavior as discussed in the work of De \textit{et al.}\cite{de2010percolation}. A detailed analysis is presented in the Supplementary Information, Section S3.

\subsection*{\raggedright{van der Waals Heterostructures}}
In order to demonstrate proof-of-principle production of a vdWH, Raman spectra were obtained at each stage in the fabrication of a FLG-MoS$_2$-WS$_2$ vdWH deposited on a glass slide. These are shown in Figure \ref{fig:3}b) along with photographs obtained from the same region of the slide at each deposition step, which are indicative of a high degree of film uniformity. The spectra, obtained from a pure FLG film (red dashed line), a FLG-MoS$_2$ heterobilayer (green dash-dot line) and a heterotrilayer of FLG-MoS$_2$-WS$_2$ vdWH (blue line), can be understood to be a linear combination of spectra from each individual material. However, subtle changes can be observed to the graphene-related modes upon deposition of the subsequent TMDC layers. In particular, the deposition of MoS$_2$ to form the FLG-MoS$_2$ heterobilayer leads to a measurable change in the ratio between the FLG $D$ and $G$ peaks, $I_D/I_G$, from 0.48 to 0.77, which recovers somewhat to 0.58 after further deposition of a WS$_2$ layer. Likewise there is a decrease in the ratio of the $G$ to $D'$ peak intensities, which indicates that the cause is a loss of $G$ band intensity. The (partial) reversibility of the change to the graphene Raman spectrum would suggest that the origin of this behavior (also observed in a FLG-MoS$_2$-FLG heterostructure, Supplementary Information, Figure S12) does not arise from damage or modification to the bottom FLG layer upon stacking of subsequent layers of TMDCs. Siimilar changes to Raman spectra of graphene layers have been observed upon electrostatic doping\cite{Das2008} and in the assembly of other heterostructures\cite{froehlicher2018} indicating that the changes to the graphene Raman spectra upon assembly of vdWHs may be explained by doping originating from charge transfer between the FLG and the TMDC.

It has been shown that a novel approach to liquid-liquid interface assembly, in which film assembly is coupled with removal of surfactant from a stabilized aqueous suspension of two dimensional material, can be used to deposit thin films and van der Waals heterostructures of 2DLMs. This generic technique may be applied to 2DLM platelets which can be held in an aqueous surfactant-stabilized suspension and to a broad spectrum of solid substrates. The effectiveness of the technique has been demonstrated through the deposition of films of FLG and TMDCs on copper foils, natively oxidized silicon wafers, and glass slides; the formation of transparent, conductive FLG films with characteristics comparable to those reported in the literature; and vdWHs of FLG, MoS$_2$, and WS$_2$. The simplicity, scalability, generic nature, broad applicability and low environmental footprint of LID opens a route for the routine use of 2DLM thin films and vdWHs in a range of applications.

\section*{\raggedright{Methods}}
\subsection*{\raggedright{Production of surfactant-stabilized aqueous 2DLM suspensions}}
Surfactant-stabilized aqueous  suspensions of 2DLMs were produced by a common approach through high shear exfoliation of a `parent' layered 3D solid in ultra-pure water (Milipore-Q, 18.2~M$\Omega\cdot$cm) in the presence of a surfactant. For the work presented here either Triton X-100 (Sigma Aldrich) or Tween 20 (Sigma Aldrich) was used, although any surfactant which meets the criterion of having a high partition coefficient into the non-aqueous phase (as discussed above) should be suitable. Shear exfoliation of the 2DLM parent material was performed using a Silverson L5M shear mixer at speeds between $2000-9000$~rpm, corresponding to shear rates of $23-104$~kHz. The resulting suspension was then centrifuged (Eppendorf 5804) to eliminate thicker and larger platelets from the suspension\cite{green2009,backes2016}. Surfactant-stabilized aqueous suspensions of graphene, MoS$_2$, MoSe$_2$, and WS$_2$ were produced from their as-received bulk parent materials (Alfa Aesar, Sigma Aldrich).

\subsection*{\raggedright{LID thin film deposition}} 
For thin film and vdWH deposition, an aqueous precursor surfactant-stabilzed suspension (approximately 2~ml) was added to a 50~ml centrifuge tube containing a `separation solvent' (approximately 22.5~ml), DCM. An emulsion of the surfactant-stabilized aqueous suspension and DCM (Fisher Scientific, 99.8+\%) was produced by shaking, using a vortex mixer (Vortex-Genie 2). The color of the emulsion was dependent on the precursor material (e.g., grey for graphene, green for MoS$_2$). The emulsion was placed in an ultrasonic bath until a clear separation between the immiscible liquids was observed. The sample was then centrifuged for 25 minutes at 11,000~rpm (corresponding to a relative centrifugal force of 16639~$g$)\cite{eppendorf}, leading to the formation of a thin continuous film at the interface between the two immiscible phases. The previously colored aqueous suspension which lay above the interface became clear and consisted only of pure water, to the limit of detection with UV-vis (Cary 5000 UV-Vis-NIR Spectrophotometer, Shimadzu UV-3600 UV-Vis-NIR Spectrophotometer) and Raman (ASEQ RM1, 532~nm, 10$\times$ objective, power $\sim$18~mW~cm$^{-2}$) spectroscopies. Below the interface the phase consisted of a solution of Triton X-100 in DCM with small aggregates of thick 2DLM platelets at the bottom of the centrifuge tube. The centrifuging step is not essential, if the centrifuge tube containing the DCM and aqueous suspension is left to settle over approximately 24 hours, the surfactant will migrate across the interface into the DCM and a film will form at the interface between the two liquids. Scaling of the process can therefore be achieved via using larger centrifuges or large tanks for gravitationally-driven assembly.

Once formed at the interface between the two immiscible liquids the film was transferred to a variety of substrates, both hydrophobic and hydrophilic, by dipping, or by other methods discussed in the Supplementary Information. Analysis using Raman spectroscopy of the thin films and heterostructures showed no evidence of the presence of the surfactant, separation solvent, or any other contaminants, indicating that the film formation process simultaneously removed the surfactant and other sources of gross contamination. There was also no evidence of surfactant when performing SEM (Zeiss Sigma 300 VP) or TEM analysis of samples. Deposition through dipping was found to be the most versatile and effective approach and is hence discussed in detail here, with the results of the other techniques reported in the Supplementary Information Section S1. 

The 2DLM platelets used for the production of films have a maximum area of $\sim$60~$\mu \mathrm{m}^2$ and a maximum lateral extent of $\sim$7~$\mu \mathrm{m}$ (a typical size distribution is shown in Supplementary Information, Figure S13). Platelet size distributions were determined by vacuum filtering an aliquot of the surfactant-stabilized aqueous suspension onto cellulose nitrate (Whatman, pore size 0.1~$\mu$m) filter membranes. The few-layer 2DLM platelets were thoroughly washed in ultrapure water and then imaged with a scanning electron microscope using backscattered electrons. The resultant images are then analyzed to determine metrics related to platelet dimensions, including the distribution of area and lateral size, using custom-developed code. Typically, $>3500$ platelets were used to determine size distributions. Full details of the semi-automated platelet characterization method will be reported elsewhere (in preparation, A.R.S., M.S., M.R.C.H.).

\subsection*{\raggedright{Film and heterostructure characterisation}}
The transmittance of thin films and vdWHs deposited on transparent substrates was measured using UV-Vis-NIR spectroscopy, using the same apparatus used to characterize the liquid phases, described above. The uniformity of deposition was measured using a CanoScan LiDE 200 flatbed scanner, following the approach of De \textit{et al.}\cite{de2009}, with transmittance calibrated using a range of neutral density filters and corrected for the background absorption of the substrate through use of a reference without the presence of the film. Raman spectroscopy was carried out using the spectrometer described above, with appropriate background subtraction. The background signal from a Triton X-100/water mixture of the same concentration as that used to form the surfactant-stabilized aqueous suspension has been subtracted from the spectrum from the suspension presented in Figure \ref{fig:2}a). Likewise, background originating from a bare glass slide has been subtracted from that of the FLG and MoS$_2$ (Figure \ref{fig:2}c)) films deposited on glass and the vdWHs presented in Figure \ref{fig:3}b). Transmission electron microscopy was undertaken on films deposited on holey carbon grids using a JEOL 3100 microscope operating at beam voltage of 200~kV. 

Film sheet resistance was measured using a home-built four probe device consisting of spring-loaded stainless steel contacts arranged in a van der Pauw geometry. A Keithley 2420 3A Source-Meter was used as a current source with voltage measured with a Keithley 2000 Multimeter and transmittance corresponding to the region probed in each four-point measurement was found using the calibrated flatbed scanner, as described above. The FLG films were annealed under inert (argon) atmosphere in a tube furnace (Carbolite CTF 12/65/550) at various fixed temperatures (100, 200 and 300~{\textdegree}C) for 24 hours in order to improve conductivity. Transmittance and sheet resistance were measured after each anneal, with the samples cooled to room temperature and under ambient atmosphere. Conductivity data were derived from sheet resistance with nominal film thickness determined by assuming 2.3\% light absorption per graphene layer\cite{Nair2008} and a turbostratic interlayer spacing\cite{kokmat2023} of $\sim 0.35$~nm.

\subsection*{\raggedright{Sustainability}}
Our technique is specifically developed for aqueous surfactant-stabilized suspensions, eliminating the need for hazardous organic solvents at this stage. It is also possible to use biodegradable surfactants to produce these suspensions. The separation solvent, DCM used to obtain the data presented in this paper, was recycled for further use using a rotary evaporator (Heidolph Laborota 4001).

\section*{\raggedright{Acknowledgements}}
The authors would like to thank Dr Lars-Olof P\aa lsson for the use of his UV-vis spectrometer. We would also like to thank Dr Alina Talmantaite and Prof Budhika Mendis for their assistance with TEM imaging. The authors would also like to thank Mr Leon Bowen and the G.J. Russell Microscopy Facility for assistance with this work. This work was funded by EPSRC Grant Number EP/T518001/1. The funder played no role in study design, data collection, analysis and interpretation of data, or the writing of this manuscript.

\section*{\raggedright{Conflict of Interest}}
All authors declare no financial or non-financial competing interests. 

\section*{\raggedright{Author Contributions}}
A.R.S., M.Z., and A.G.M.M. performed sample deposition. A.R.S. optimised the deposition, designed the automated dipping machine, performed conductivity measurements, and took transmission scans. M.R.C.H. performed Raman measurements. A.R.S. and M.R.C.H. analysed the data. A.R.S., M.R.C.H., and M.S. wrote the initial draft of the manuscript and A.R.S. prepared figures 1-3. All authors edited and approved the manuscript. M.R.C.H. and M.S. conceived and supervised the project.

\section*{\raggedright{Data Availability}}
The authors declare that the data supporting the findings are available within the paper and its supplementary information. The corresponding authors can also provide data upon request.

\bibliographystyle{unsrt}

\bibliography{LID_paper}

\begin{thebibliography}{10}

\bibitem{Schwierz2010}
Frank Schwierz.
\newblock {Graphene transistors.}
\newblock {\em Nature Nanotechnology}, 5(7):487--96, 2010.

\bibitem{Schwierz2011}
Frank Schwierz.
\newblock {Nanoelectronics: Flat transistors get off the ground.}
\newblock {\em Nature Nanotechnology}, 6(3):135--6, 2011.

\bibitem{Radisavljevic2011}
B~Radisavljevic, A~Radenovic, J~Brivio, V~Giacometti, and A~Kis.
\newblock {Single-layer MoS$_2$ transistors.}
\newblock {\em Nature Nanotechnology}, 6(3):147--150, 2011.

\bibitem{Hopf2014}
T~Hopf, K~V Vassilevski, E~Escobedo-Cousin, P~J King, N~G Wright, A~G O'Neill,
  A~B Horsfall, J~P Goss, G~H Wells, and M~R~C Hunt.
\newblock {Dirac point and transconductance of top-gated graphene field-effect
  transistors operating at elevated temperature.}
\newblock {\em Journal of Applied Physics}, 116(15):154504, 2014.

\bibitem{Rhee2024}
Dongjoon Rhee, Deep Jariwala, Jeong~Ho Cho, and Joohoon Kang.
\newblock {Solution-processed 2D van der Waals networks: Fabrication
  strategies, properties, and scalable device applications.}
\newblock {\em Applied Physics Reviews}, 11(2), 2024.

\bibitem{Wang2008}
Xuan Wang, Linjie Zhi, and Klaus Müllen.
\newblock {Transparent, Conductive Graphene Electrodes for Dye-Sensitized Solar
  Cells.}
\newblock {\em Nano Letters}, 8(1):323--327, 2008.

\bibitem{ricciardulli2020}
Antonio~Gaetano Ricciardulli and Paul~WM Blom.
\newblock {Solution-processable 2D materials applied in light-emitting diodes
  and solar cells.}
\newblock {\em Advanced Materials Technologies}, 5(8):1900972, 2020.

\bibitem{Varrla2015}
Eswaraiah Varrla, Claudia Backes, Keith~R Paton, Andrew Harvey, Zahra
  Gholamvand, Joe McCauley, and Jonathan~N Coleman.
\newblock {Large-Scale Production of Size-Controlled MoS$_2$ Nanosheets by
  Shear Exfoliation.}
\newblock {\em Chemistry of Materials}, 27(3):1129--1139, 2015.

\bibitem{Li2018}
Tian Li, Andrea~D Pickel, Yonggang Yao, Yanan Chen, Yuqiang Zeng, Steven~D
  Lacey, Yiju Li, Yilin Wang, Jiaqi Dai, Yanbin Wang, Bao Yang, Michael~S
  Fuhrer, Amy Marconnet, Chris Dames, Dennis~H Drew, and Liangbing Hu.
\newblock {Thermoelectric properties and performance of flexible reduced
  graphene oxide films up to 3,000 K.}
\newblock {\em Nature Energy}, 3(2):148--156, 2018.

\bibitem{De2010}
Sukanta De, Paul~J King, Mustafa Lotya, Arlene O'Neill, Evelyn~M Doherty, Yenny
  Hernandez, Georg~S Duesberg, and Jonathan~N Coleman.
\newblock {Flexible, Transparent, Conducting Films of Randomly Stacked Graphene
  from Surfactant-Stabilized, Oxide-Free Graphene Dispersions.}
\newblock {\em Small}, 6(3):458--464, 2010.

\bibitem{Matkovic2016}
Aleksandar Matković, Ivana Milošević, Marijana Milićević, Tijana
  Tomašević-Ilić, Jelena Pešić, Milenko Musić, Marko Spasenović, Djordje
  Jovanović, Borislav Vasić, Christopher Deeks, Radmila Panajotović,
  Milivoj~R Belić, and Radoš Gajić.
\newblock {Enhanced sheet conductivity of Langmuir–Blodgett assembled
  graphene thin films by chemical doping.}
\newblock {\em 2D Materials}, 3(1):015002, 2016.

\bibitem{Woltornist2017}
Steven~J Woltornist, Deepthi Varghese, Daniel Massucci, Zhen Cao, Andrey~V
  Dobrynin, and Douglas~H Adamson.
\newblock {Controlled 3D Assembly of Graphene Sheets to Build Conductive,
  Chemically Selective and Shape-Responsive Materials.}
\newblock {\em Advanced Materials}, 29(18):1604947, 2017.

\bibitem{carey2024}
Tian Carey, Jack Maughan, Luke Doolan, Eoin Caffrey, James Garcia, Shixin Liu,
  Harneet Kaur, Cansu Ilhan, Shayan Seyedin, and Jonathan~N Coleman.
\newblock {Knot Architecture for Biocompatible and Semiconducting 2D Electronic
  Fiber Transistors.}
\newblock {\em Small Methods}, 2024.
\newblock 2301654.

\bibitem{Geim2013}
Andre~K Geim and Irina~V Grigorieva.
\newblock {Van der Waals heterostructures.}
\newblock {\em Nature}, 499(7459):419--425, 2013.

\bibitem{Castellanos-Gomez2014}
Andres Castellanos-Gomez, Michele Buscema, Rianda Molenaar, Vibhor Singh,
  Laurens Janssen, Herre S.~J. van~der Zant, and Gary~A. Steele.
\newblock {Deterministic transfer of two-dimensional materials by all-dry
  viscoelastic stamping.}
\newblock {\em 2D Materials}, 1(1), 2014.

\bibitem{kim2022}
Jihyun Kim, Dongjoon Rhee, Okin Song, Miju Kim, Yong~Hyun Kwon, Dong~Un Lim,
  In~Soo Kim, Vlastimil Maz{\'a}nek, Lukas Valdman, Zden{\v{e}}k Sofer, et~al.
\newblock {All-solution-processed van der Waals heterostructures for
  wafer-scale electronics.}
\newblock {\em Advanced Materials}, 34(12):2106110, 2022.

\bibitem{zhang2020}
Jing Zhang, Fan Wang, Vivek~B Shenoy, Ming Tang, and Jun Lou.
\newblock {Towards controlled synthesis of 2D crystals by chemical vapor
  deposition (CVD)}.
\newblock {\em Materials Today}, 40:132--139, 2020.

\bibitem{Woltornist2013}
Steven~J Woltornist, Andrew~J Oyer, Jan-Michael~Y Carrillo, Andrey~V Dobrynin,
  and Douglas~H Adamson.
\newblock {Conductive Thin Films of Pristine Graphene by Solvent Interface
  Trapping.}
\newblock {\em ACS Nano}, 7(8):7062--7066, 2013.

\bibitem{Clark2017}
R.~M. Clark, K.~J. Berean, B.~J. Carey, N.~Pillai, T.~Daeneke, I.~S. Cole,
  K.~Latham, and K.~Kalantar-zadeh.
\newblock {Patterned films from exfoliated two-dimensional transition metal
  dichalcogenides assembled at a liquid–liquid interface.}
\newblock {\em Journal of Materials Chemistry C}, 5(28):6937--6944, 2017.

\bibitem{Yu2015}
X.~Yu, M.~S. Prevot, N.~Guijarro, and K.~Sivula.
\newblock {Self-assembled 2D WSe$_2$ thin films for photoelectrochemical
  hydrogen production.}
\newblock {\em Nat Commun}, 6:7596, 2015.

\bibitem{Neilson2020}
J.~Neilson, M.~P. Avery, and B.~Derby.
\newblock {Tiled Monolayer Films of 2D Molybdenum Disulfide Nanoflakes
  Assembled at Liquid/Liquid Interfaces.}
\newblock {\em ACS Appl Mater Interfaces}, 12(22):25125--25134, 2020.

\bibitem{cassidy2025}
Oran Cassidy, Kevin Synnatschke, Jose~M Munuera, Cian Gabbett, Tian Carey, Luke
  Doolan, Eoin Caffrey, and Jonathan~N Coleman.
\newblock {Layer-by-layer assembly yields thin graphene films with near
  theoretical conductivity.}
\newblock {\em npj 2D Materials and Applications}, 9(1):2, 2025.

\bibitem{paton2014}
Keith~R Paton, Eswaraiah Varrla, Claudia Backes, Ronan~J Smith, Umar Khan,
  Arlene O’Neill, Conor Boland, Mustafa Lotya, Oana~M Istrate, Paul King,
  et~al.
\newblock {Scalable production of large quantities of defect-free few-layer
  graphene by shear exfoliation in liquids.}
\newblock {\em Nature Materials}, 13(6):624--630, 2014.

\bibitem{bigcentrifuges}
Beckman Coulter.
\newblock {High-Performance \& High-Capacity Centrifuges: Product Catalog.}
\newblock
  \url{https://uk.vwr.com/assetsvc/asset/en_GB/id/39706114/contents/beckman-coulter-high-performance-centrifuge.pdf}.

\bibitem{shen2015liquid}
Jianfeng Shen, Yongmin He, Jingjie Wu, Caitian Gao, Kunttal Keyshar, Xiang
  Zhang, Yingchao Yang, Mingxin Ye, Robert Vajtai, Jun Lou, et~al.
\newblock {Liquid phase exfoliation of two-dimensional materials by directly
  probing and matching surface tension components}.
\newblock {\em Nano Letters}, 15(8):5449--5454, 2015.

\bibitem{Lee2010}
C.~Lee, H.~Yan, L.~E. Brus, T.~F. Heinz, J.~Hone, and S.~Ryu.
\newblock {Anomalous lattice vibrations of single- and few-layer MoS$_2$.}
\newblock {\em ACS Nano}, 4(5):2695--700, 2010.

\bibitem{Mak2010}
Kin Mak, Changgu Lee, James Hone, Jie Shan, and Tony Heinz.
\newblock {Atomically Thin MoS$_2$: A New Direct-Gap Semiconductor}.
\newblock {\em Physical Review Letters}, 105(13):2--5, 2010.

\bibitem{Nair2008}
R.~R. Nair, P.~Blake, A.~N. Grigorenko, K.~S. Novoselov, T.~J. Booth,
  T.~Stauber, N.~M.~R. Peres, and A.~K. Geim.
\newblock {Fine Structure Constant Defines Visual Transparency of Graphene.}
\newblock {\em Science}, 320, 2008.

\bibitem{liu2013electrochemically}
Jinzhang Liu, Marco Notarianni, Geoffrey Will, Vincent~Tiing Tiong, Hongxia
  Wang, and Nunzio Motta.
\newblock {Electrochemically exfoliated graphene for electrode films: effect of
  graphene flake thickness on the sheet resistance and capacitive properties}.
\newblock {\em Langmuir}, 29(43):13307--13314, 2013.

\bibitem{de2010percolation}
Sukanta De, Paul~J King, Philip~E Lyons, Umar Khan, and Jonathan~N Coleman.
\newblock {Size effects and the problem with percolation in nanostructured
  transparent conductors.}
\newblock {\em ACS Nano}, 4(12):7064--7072, 2010.

\bibitem{Das2008}
A.~Das, S.~Pisana, B.~Chakraborty, S.~Piscanec, S.~K. Saha, U.~V. Waghmare,
  K.~S. Novoselov, H.~R. Krishnamurthy, A.~K. Geim, A.~C. Ferrari, and A.~K.
  Sood.
\newblock {Monitoring dopants by Raman scattering in an electrochemically
  top-gated graphene transistor.}
\newblock {\em Nat Nanotechnol}, 3(4):210--5, 2008.

\bibitem{froehlicher2018}
Guillaume Froehlicher, Etienne Lorchat, and St{\'e}phane Berciaud.
\newblock {Charge versus energy transfer in atomically thin graphene-transition
  metal dichalcogenide van der Waals heterostructures.}
\newblock {\em Physical Review X}, 8(1):011007, 2018.

\bibitem{green2009}
Alexander~A Green and Mark~C Hersam.
\newblock {Solution phase production of graphene with controlled thickness via
  density differentiation.}
\newblock {\em Nano Letters}, 9(12):4031--4036, 2009.

\bibitem{backes2016}
Claudia Backes, Beata~M Szyd{\l}owska, Andrew Harvey, Shengjun Yuan, Victor
  Vega-Mayoral, Ben~R Davies, Pei-liang Zhao, Damien Hanlon, Elton~JG Santos,
  Mikhail~I Katsnelson, et~al.
\newblock {Production of highly monolayer enriched dispersions of
  liquid-exfoliated nanosheets by liquid cascade centrifugation.}
\newblock {\em ACS Nano}, 10(1):1589--1601, 2016.

\bibitem{eppendorf}
{Eppendorf SE}.
\newblock {\em {Centrifuge 5804/5804 R Centrifuge 5810/5810 R Original
  Operating Instructions.}}, first edition, 2021.

\bibitem{de2009}
Sukanta De, Philip~E Lyons, Sophie Sorel, Evelyn~M Doherty, Paul~J King,
  Werner~J Blau, Peter~N Nirmalraj, John~J Boland, Vittorio Scardaci, Jerome
  Joimel, et~al.
\newblock {Transparent, flexible, and highly conductive thin films based on
  polymer-nanotube composites.}
\newblock {\em ACS Nano}, 3(3):714--720, 2009.

\bibitem{kokmat2023}
Phurida Kokmat, Piyaporn Surinlert, and Akkawat Ruammaitree.
\newblock {Growth of High-Purity and High-Quality Turbostratic Graphene with
  Different Interlayer Spacings.}
\newblock {\em ACS Omega}, 8(4):4010--4018, 2023.

\end{thebibliography}


\begin{thebibliography}{11}%
\makeatletter
\providecommand \@ifxundefined [1]{%
 \@ifx{#1\undefined}
}%
\providecommand \@ifnum [1]{%
 \ifnum #1\expandafter \@firstoftwo
 \else \expandafter \@secondoftwo
 \fi
}%
\providecommand \@ifx [1]{%
 \ifx #1\expandafter \@firstoftwo
 \else \expandafter \@secondoftwo
 \fi
}%
\providecommand \natexlab [1]{#1}%
\providecommand \enquote  [1]{``#1''}%
\providecommand \bibnamefont  [1]{#1}%
\providecommand \bibfnamefont [1]{#1}%
\providecommand \citenamefont [1]{#1}%
\providecommand \href@noop [0]{\@secondoftwo}%
\providecommand \href [0]{\begingroup \@sanitize@url \@href}%
\providecommand \@href[1]{\@@startlink{#1}\@@href}%
\providecommand \@@href[1]{\endgroup#1\@@endlink}%
\providecommand \@sanitize@url [0]{\catcode `\\12\catcode `\$12\catcode
  `\&12\catcode `\#12\catcode `\^12\catcode `\_12\catcode `\%12\relax}%
\providecommand \@@startlink[1]{}%
\providecommand \@@endlink[0]{}%
\providecommand \url  [0]{\begingroup\@sanitize@url \@url }%
\providecommand \@url [1]{\endgroup\@href {#1}{\urlprefix }}%
\providecommand \urlprefix  [0]{URL }%
\providecommand \Eprint [0]{\href }%
\providecommand \doibase [0]{https://doi.org/}%
\providecommand \selectlanguage [0]{\@gobble}%
\providecommand \bibinfo  [0]{\@secondoftwo}%
\providecommand \bibfield  [0]{\@secondoftwo}%
\providecommand \translation [1]{[#1]}%
\providecommand \BibitemOpen [0]{}%
\providecommand \bibitemStop [0]{}%
\providecommand \bibitemNoStop [0]{.\EOS\space}%
\providecommand \EOS [0]{\spacefactor3000\relax}%
\providecommand \BibitemShut  [1]{\csname bibitem#1\endcsname}%
\let\auto@bib@innerbib\@empty
\bibitem [{\citenamefont {De}\ \emph {et~al.}(2010{\natexlab{a}})\citenamefont
  {De}, \citenamefont {King}, \citenamefont {Lotya}, \citenamefont {O'Neill},
  \citenamefont {Doherty}, \citenamefont {Hernandez}, \citenamefont
  {Duesberg},\ and\ \citenamefont {Coleman}}]{De2010}%
  \BibitemOpen
  \bibfield  {author} {\bibinfo {author} {\bibfnamefont {S.}~\bibnamefont
  {De}}, \bibinfo {author} {\bibfnamefont {P.~J.}\ \bibnamefont {King}},
  \bibinfo {author} {\bibfnamefont {M.}~\bibnamefont {Lotya}}, \bibinfo
  {author} {\bibfnamefont {A.}~\bibnamefont {O'Neill}}, \bibinfo {author}
  {\bibfnamefont {E.~M.}\ \bibnamefont {Doherty}}, \bibinfo {author}
  {\bibfnamefont {Y.}~\bibnamefont {Hernandez}}, \bibinfo {author}
  {\bibfnamefont {G.~S.}\ \bibnamefont {Duesberg}},\ and\ \bibinfo {author}
  {\bibfnamefont {J.~N.}\ \bibnamefont {Coleman}},\ }\bibfield  {title}
  {\bibinfo {title} {{Flexible, Transparent, Conducting Films of Randomly
  Stacked Graphene from Surfactant-Stabilized, Oxide-Free Graphene
  Dispersions}},\ }\href@noop {} {\bibfield  {journal} {\bibinfo  {journal}
  {Small}\ }\textbf {\bibinfo {volume} {6}},\ \bibinfo {pages} {458} (\bibinfo
  {year} {2010}{\natexlab{a}})}\BibitemShut {NoStop}%
\bibitem [{\citenamefont {De}\ \emph {et~al.}(2010{\natexlab{b}})\citenamefont
  {De}, \citenamefont {King}, \citenamefont {Lyons}, \citenamefont {Khan},\
  and\ \citenamefont {Coleman}}]{de2010percolation}%
  \BibitemOpen
  \bibfield  {author} {\bibinfo {author} {\bibfnamefont {S.}~\bibnamefont
  {De}}, \bibinfo {author} {\bibfnamefont {P.~J.}\ \bibnamefont {King}},
  \bibinfo {author} {\bibfnamefont {P.~E.}\ \bibnamefont {Lyons}}, \bibinfo
  {author} {\bibfnamefont {U.}~\bibnamefont {Khan}},\ and\ \bibinfo {author}
  {\bibfnamefont {J.~N.}\ \bibnamefont {Coleman}},\ }\bibfield  {title}
  {\bibinfo {title} {Size effects and the problem with percolation in
  nanostructured transparent conductors},\ }\href@noop {} {\bibfield  {journal}
  {\bibinfo  {journal} {ACS Nano}\ }\textbf {\bibinfo {volume} {4}},\ \bibinfo
  {pages} {7064} (\bibinfo {year} {2010}{\natexlab{b}})}\BibitemShut {NoStop}%
\bibitem [{\citenamefont {Mojtabavi}\ \emph {et~al.}(2021)\citenamefont
  {Mojtabavi}, \citenamefont {VahidMohammadi}, \citenamefont {Ganeshan},
  \citenamefont {Hejazi}, \citenamefont {Shahbazmohamadi}, \citenamefont {Kar},
  \citenamefont {Van~Duin},\ and\ \citenamefont {Wanunu}}]{mojtabavi2021wafer}%
  \BibitemOpen
  \bibfield  {author} {\bibinfo {author} {\bibfnamefont {M.}~\bibnamefont
  {Mojtabavi}}, \bibinfo {author} {\bibfnamefont {A.}~\bibnamefont
  {VahidMohammadi}}, \bibinfo {author} {\bibfnamefont {K.}~\bibnamefont
  {Ganeshan}}, \bibinfo {author} {\bibfnamefont {D.}~\bibnamefont {Hejazi}},
  \bibinfo {author} {\bibfnamefont {S.}~\bibnamefont {Shahbazmohamadi}},
  \bibinfo {author} {\bibfnamefont {S.}~\bibnamefont {Kar}}, \bibinfo {author}
  {\bibfnamefont {A.~C.}\ \bibnamefont {Van~Duin}},\ and\ \bibinfo {author}
  {\bibfnamefont {M.}~\bibnamefont {Wanunu}},\ }\bibfield  {title} {\bibinfo
  {title} {{Wafer-scale lateral self-assembly of mosaic {Ti$_3$C$_2$T$_x$
  MXene} monolayer films}},\ }\href@noop {} {\bibfield  {journal} {\bibinfo
  {journal} {ACS Nano}\ }\textbf {\bibinfo {volume} {15}},\ \bibinfo {pages}
  {625} (\bibinfo {year} {2021})}\BibitemShut {NoStop}%
\bibitem [{\citenamefont {Shi}\ \emph {et~al.}(2024)\citenamefont {Shi},
  \citenamefont {Li}, \citenamefont {Tsunematsu}, \citenamefont {Ozeki},
  \citenamefont {Kano}, \citenamefont {Yamamoto}, \citenamefont {Kobayashi},
  \citenamefont {Abe}, \citenamefont {Chen},\ and\ \citenamefont
  {Osada}}]{shi2024ultrafast}%
  \BibitemOpen
  \bibfield  {author} {\bibinfo {author} {\bibfnamefont {Y.}~\bibnamefont
  {Shi}}, \bibinfo {author} {\bibfnamefont {H.}~\bibnamefont {Li}}, \bibinfo
  {author} {\bibfnamefont {H.}~\bibnamefont {Tsunematsu}}, \bibinfo {author}
  {\bibfnamefont {H.}~\bibnamefont {Ozeki}}, \bibinfo {author} {\bibfnamefont
  {K.}~\bibnamefont {Kano}}, \bibinfo {author} {\bibfnamefont {E.}~\bibnamefont
  {Yamamoto}}, \bibinfo {author} {\bibfnamefont {M.}~\bibnamefont {Kobayashi}},
  \bibinfo {author} {\bibfnamefont {H.}~\bibnamefont {Abe}}, \bibinfo {author}
  {\bibfnamefont {C.-W.}\ \bibnamefont {Chen}},\ and\ \bibinfo {author}
  {\bibfnamefont {M.}~\bibnamefont {Osada}},\ }\bibfield  {title} {\bibinfo
  {title} {{Ultrafast 2D Nanosheet Assembly via Spontaneous Spreading
  Phenomenon}},\ }\href@noop {} {\bibfield  {journal} {\bibinfo  {journal}
  {Small}\ ,\ \bibinfo {pages} {2403915}} (\bibinfo {year} {2024})}\BibitemShut
  {NoStop}%
\bibitem [{\citenamefont {Li}\ \emph {et~al.}(2014)\citenamefont {Li},
  \citenamefont {Naiini}, \citenamefont {Vaziri}, \citenamefont {Lemme},\ and\
  \citenamefont {{\"O}stling}}]{li2014inkjet}%
  \BibitemOpen
  \bibfield  {author} {\bibinfo {author} {\bibfnamefont {J.}~\bibnamefont
  {Li}}, \bibinfo {author} {\bibfnamefont {M.~M.}\ \bibnamefont {Naiini}},
  \bibinfo {author} {\bibfnamefont {S.}~\bibnamefont {Vaziri}}, \bibinfo
  {author} {\bibfnamefont {M.~C.}\ \bibnamefont {Lemme}},\ and\ \bibinfo
  {author} {\bibfnamefont {M.}~\bibnamefont {{\"O}stling}},\ }\bibfield
  {title} {\bibinfo {title} {{Inkjet printing of {MoS$_2$}}},\ }\href@noop {}
  {\bibfield  {journal} {\bibinfo  {journal} {Advanced Functional Materials}\
  }\textbf {\bibinfo {volume} {24}},\ \bibinfo {pages} {6524} (\bibinfo {year}
  {2014})}\BibitemShut {NoStop}%
\bibitem [{\citenamefont {Kim}\ \emph {et~al.}(2021)\citenamefont {Kim},
  \citenamefont {Kim}, \citenamefont {Cho}, \citenamefont {Choi}, \citenamefont
  {Jung}, \citenamefont {Cho}, \citenamefont {Whang},\ and\ \citenamefont
  {Kang}}]{kim2021solution}%
  \BibitemOpen
  \bibfield  {author} {\bibinfo {author} {\bibfnamefont {J.}~\bibnamefont
  {Kim}}, \bibinfo {author} {\bibfnamefont {S.}~\bibnamefont {Kim}}, \bibinfo
  {author} {\bibfnamefont {Y.~S.}\ \bibnamefont {Cho}}, \bibinfo {author}
  {\bibfnamefont {M.}~\bibnamefont {Choi}}, \bibinfo {author} {\bibfnamefont
  {S.-H.}\ \bibnamefont {Jung}}, \bibinfo {author} {\bibfnamefont {J.~H.}\
  \bibnamefont {Cho}}, \bibinfo {author} {\bibfnamefont {D.}~\bibnamefont
  {Whang}},\ and\ \bibinfo {author} {\bibfnamefont {J.}~\bibnamefont {Kang}},\
  }\bibfield  {title} {\bibinfo {title} {{Solution-processed {MoS$_2$} film
  with functional interfaces via precursor-assisted chemical welding}},\
  }\href@noop {} {\bibfield  {journal} {\bibinfo  {journal} {ACS Applied
  Materials \& Interfaces}\ }\textbf {\bibinfo {volume} {13}},\ \bibinfo
  {pages} {12221} (\bibinfo {year} {2021})}\BibitemShut {NoStop}%
\bibitem [{\citenamefont {Kim}\ \emph {et~al.}(2022)\citenamefont {Kim},
  \citenamefont {Rhee}, \citenamefont {Song}, \citenamefont {Kim},
  \citenamefont {Kwon}, \citenamefont {Lim}, \citenamefont {Kim}, \citenamefont
  {Maz{\'a}nek}, \citenamefont {Valdman}, \citenamefont {Sofer} \emph
  {et~al.}}]{kim2022all}%
  \BibitemOpen
  \bibfield  {author} {\bibinfo {author} {\bibfnamefont {J.}~\bibnamefont
  {Kim}}, \bibinfo {author} {\bibfnamefont {D.}~\bibnamefont {Rhee}}, \bibinfo
  {author} {\bibfnamefont {O.}~\bibnamefont {Song}}, \bibinfo {author}
  {\bibfnamefont {M.}~\bibnamefont {Kim}}, \bibinfo {author} {\bibfnamefont
  {Y.~H.}\ \bibnamefont {Kwon}}, \bibinfo {author} {\bibfnamefont {D.~U.}\
  \bibnamefont {Lim}}, \bibinfo {author} {\bibfnamefont {I.~S.}\ \bibnamefont
  {Kim}}, \bibinfo {author} {\bibfnamefont {V.}~\bibnamefont {Maz{\'a}nek}},
  \bibinfo {author} {\bibfnamefont {L.}~\bibnamefont {Valdman}}, \bibinfo
  {author} {\bibfnamefont {Z.}~\bibnamefont {Sofer}}, \emph {et~al.},\
  }\bibfield  {title} {\bibinfo {title} {All-solution-processed van der waals
  heterostructures for wafer-scale electronics},\ }\href@noop {} {\bibfield
  {journal} {\bibinfo  {journal} {Advanced Materials}\ }\textbf {\bibinfo
  {volume} {34}},\ \bibinfo {pages} {2106110} (\bibinfo {year}
  {2022})}\BibitemShut {NoStop}%
\bibitem [{\citenamefont {Cunningham}\ \emph {et~al.}(2013)\citenamefont
  {Cunningham}, \citenamefont {Khan}, \citenamefont {Backes}, \citenamefont
  {Hanlon}, \citenamefont {McCloskey}, \citenamefont {Donegan},\ and\
  \citenamefont {Coleman}}]{cunningham2013photoconductivity}%
  \BibitemOpen
  \bibfield  {author} {\bibinfo {author} {\bibfnamefont {G.}~\bibnamefont
  {Cunningham}}, \bibinfo {author} {\bibfnamefont {U.}~\bibnamefont {Khan}},
  \bibinfo {author} {\bibfnamefont {C.}~\bibnamefont {Backes}}, \bibinfo
  {author} {\bibfnamefont {D.}~\bibnamefont {Hanlon}}, \bibinfo {author}
  {\bibfnamefont {D.}~\bibnamefont {McCloskey}}, \bibinfo {author}
  {\bibfnamefont {J.~F.}\ \bibnamefont {Donegan}},\ and\ \bibinfo {author}
  {\bibfnamefont {J.~N.}\ \bibnamefont {Coleman}},\ }\bibfield  {title}
  {\bibinfo {title} {{Photoconductivity of solution-processed {MoS$_2$}
  films}},\ }\href@noop {} {\bibfield  {journal} {\bibinfo  {journal} {Journal
  of Materials Chemistry C}\ }\textbf {\bibinfo {volume} {1}},\ \bibinfo
  {pages} {6899} (\bibinfo {year} {2013})}\BibitemShut {NoStop}%
\bibitem [{\citenamefont {Abdolhosseinzadeh}\ \emph {et~al.}(2022)\citenamefont
  {Abdolhosseinzadeh}, \citenamefont {Zhang}, \citenamefont {Schneider},
  \citenamefont {Shakoorioskooie}, \citenamefont {N{\"u}esch},\ and\
  \citenamefont {Heier}}]{abdolhosseinzadeh2022universal}%
  \BibitemOpen
  \bibfield  {author} {\bibinfo {author} {\bibfnamefont {S.}~\bibnamefont
  {Abdolhosseinzadeh}}, \bibinfo {author} {\bibfnamefont {C.}~\bibnamefont
  {Zhang}}, \bibinfo {author} {\bibfnamefont {R.}~\bibnamefont {Schneider}},
  \bibinfo {author} {\bibfnamefont {M.}~\bibnamefont {Shakoorioskooie}},
  \bibinfo {author} {\bibfnamefont {F.}~\bibnamefont {N{\"u}esch}},\ and\
  \bibinfo {author} {\bibfnamefont {J.}~\bibnamefont {Heier}},\ }\bibfield
  {title} {\bibinfo {title} {{A universal approach for room-temperature
  printing and coating of {2D} materials}},\ }\href@noop {} {\bibfield
  {journal} {\bibinfo  {journal} {Advanced Materials}\ }\textbf {\bibinfo
  {volume} {34}},\ \bibinfo {pages} {2103660} (\bibinfo {year}
  {2022})}\BibitemShut {NoStop}%
\bibitem [{\citenamefont {Higgins}\ \emph {et~al.}(2019)\citenamefont
  {Higgins}, \citenamefont {Finn}, \citenamefont {Matthiesen}, \citenamefont
  {Grieger}, \citenamefont {Synnatschke}, \citenamefont {Brohmann},
  \citenamefont {Rother}, \citenamefont {Backes},\ and\ \citenamefont
  {Zaumseil}}]{higgins2019electrolyte}%
  \BibitemOpen
  \bibfield  {author} {\bibinfo {author} {\bibfnamefont {T.~M.}\ \bibnamefont
  {Higgins}}, \bibinfo {author} {\bibfnamefont {S.}~\bibnamefont {Finn}},
  \bibinfo {author} {\bibfnamefont {M.}~\bibnamefont {Matthiesen}}, \bibinfo
  {author} {\bibfnamefont {S.}~\bibnamefont {Grieger}}, \bibinfo {author}
  {\bibfnamefont {K.}~\bibnamefont {Synnatschke}}, \bibinfo {author}
  {\bibfnamefont {M.}~\bibnamefont {Brohmann}}, \bibinfo {author}
  {\bibfnamefont {M.}~\bibnamefont {Rother}}, \bibinfo {author} {\bibfnamefont
  {C.}~\bibnamefont {Backes}},\ and\ \bibinfo {author} {\bibfnamefont
  {J.}~\bibnamefont {Zaumseil}},\ }\bibfield  {title} {\bibinfo {title}
  {{Electrolyte-gated n-type transistors produced from aqueous inks of {WS$_2$}
  nanosheets}},\ }\href@noop {} {\bibfield  {journal} {\bibinfo  {journal}
  {Advanced Functional Materials}\ }\textbf {\bibinfo {volume} {29}},\ \bibinfo
  {pages} {1804387} (\bibinfo {year} {2019})}\BibitemShut {NoStop}%
\bibitem [{\citenamefont {Zou}\ \emph {et~al.}(2023)\citenamefont {Zou},
  \citenamefont {Kim}, \citenamefont {Kim}, \citenamefont {Liu}, \citenamefont
  {Choi}, \citenamefont {Jung}, \citenamefont {Zhu}, \citenamefont {You},
  \citenamefont {Reo}, \citenamefont {Lee} \emph {et~al.}}]{zou2023high}%
  \BibitemOpen
  \bibfield  {author} {\bibinfo {author} {\bibfnamefont {T.}~\bibnamefont
  {Zou}}, \bibinfo {author} {\bibfnamefont {H.-J.}\ \bibnamefont {Kim}},
  \bibinfo {author} {\bibfnamefont {S.}~\bibnamefont {Kim}}, \bibinfo {author}
  {\bibfnamefont {A.}~\bibnamefont {Liu}}, \bibinfo {author} {\bibfnamefont
  {M.-Y.}\ \bibnamefont {Choi}}, \bibinfo {author} {\bibfnamefont
  {H.}~\bibnamefont {Jung}}, \bibinfo {author} {\bibfnamefont {H.}~\bibnamefont
  {Zhu}}, \bibinfo {author} {\bibfnamefont {I.}~\bibnamefont {You}}, \bibinfo
  {author} {\bibfnamefont {Y.}~\bibnamefont {Reo}}, \bibinfo {author}
  {\bibfnamefont {W.-J.}\ \bibnamefont {Lee}}, \emph {et~al.},\ }\bibfield
  {title} {\bibinfo {title} {{High-performance solution-processed {2D p-type
  WSe$_2$} transistors and circuits through molecular doping}},\ }\href@noop {}
  {\bibfield  {journal} {\bibinfo  {journal} {Advanced Materials}\ }\textbf
  {\bibinfo {volume} {35}},\ \bibinfo {pages} {2208934} (\bibinfo {year}
  {2023})}\BibitemShut {NoStop}%
\end{thebibliography}%

\end{document}


\pagestyle{fancy}
\begin{flushleft}
    \title{Supporting Information\\
    \normalsize Novel Liquid-Liquid Interface Deposition Method for Thin Films of Two-Dimensional Solids}
    \author{\small Amy R. Smith*, Muhammad Zulqurnain$^{\dag}$, Angus G.M. Mathieson$^{\dag}$, Marek Szablewski*, Michael R.C. Hunt*}
    \affiliation{\normalfont Email Addresses: amy.r.smith@durham.ac.uk, marek.szablewski@durham.ac.uk, m.r.c.hunt@durham.ac.uk \\
    $^{\dag}$These authors contributed equally to this work.}
    \maketitle
\end{flushleft}

\pagestyle{fancy}

\section{\raggedright{Alternative Deposition Methods}}
\begin{figure}
    \centering
    \includegraphics[width=\linewidth]{fig_S1.pdf}
    \caption{\textbf{Schematics for alternative methods of film deposition:} a)~Schematic of the horizontal dipping method, showing the substrate on the base of a trough, submerged in separation solvent and the contents of the centrifuge tube containing the 2DLM platelet film. The substrate can be lifted out of the trough through the 2DLM platelet film, depositing the film onto the substrate. b)~Pouring method of deposition. The contents centrifuge tube containing the 2DLM platelet film is gently poured over the substrate. The separation solvent evaporates, leaving the 2DLM platelet film in contact with the substrate.}
    \label{fig:Deposition_diagrams}
\end{figure}

In addition to film and heterostructure deposition through dipping, an alternate horizontal dipping approach has been used for small substrates which can be immersed in limited volumes of liquid, a schematic of which is presented in Figure \ref{fig:Deposition_diagrams}a). A plastic trough was partially filled with the organic separation solvent with a suitable substrate placed horizontally on the base of the trough. The contents of the centrifuge tube containing the two-dimensional layered material (2DLM) film assembled at the liquid-liquid interface, as described in the main paper, were carefully added, forming a droplet of water at the center of the trough with the 2DLM platelet film at its interface with the organic solvent, such that there was incomplete coverage of the separation solvent by the aqueous phase and the interfacial film. The substrate was then withdrawn upwards from the trough through the liquid-liquid interface in a manner equivalent to Langmuir-Schaeffer deposition, depositing the film on the substrate. By careful choice of the volume of suspension used in this process and appropriate manipulation of the substrate it is possible for area-selective film deposition to be achieved, which is of particular value when depositing van der Waals heterostructures (vdWHs) for device applications, as specific regions of each layer may be left exposed to enable electrical contact to be made.\\

\begin{figure}[h]
    \centering
    \includegraphics[width=0.75\linewidth]{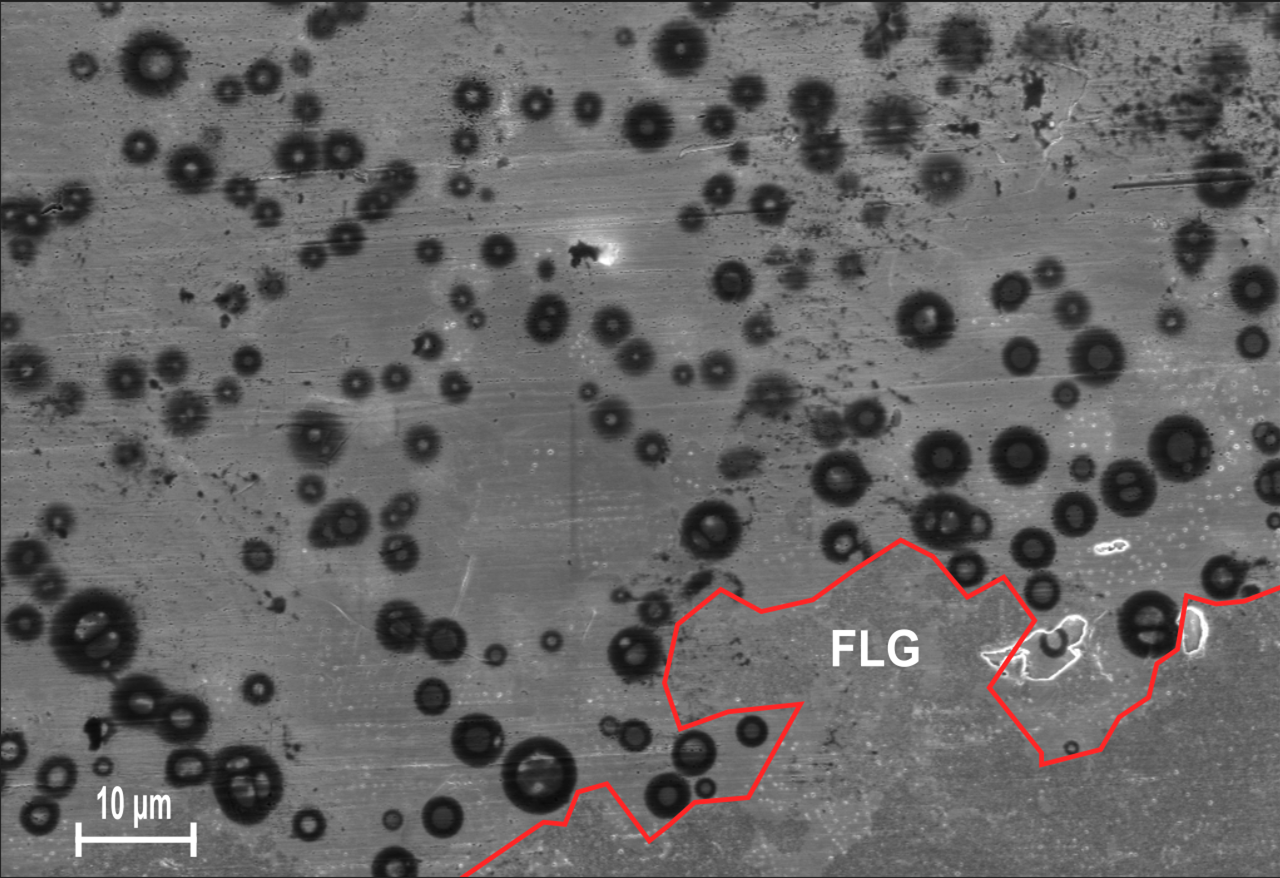}
    \caption{\textbf{SEM image showing residual surfactant.} SEM image of a few-layer graphene deposited on copper by pouring, the few-layer graphene film can be seen in the bottom right of the image, the rest of the image is covered in `coffee rings' due to residual surfactant.}
    \label{fig:coffee_rings}
\end{figure}

2DLM platelet films assembled at the liquid-liquid interface may also be transferred to almost any substrate by careful pouring of the contents of the centrifuge tube onto a chosen substrate, as shown in Figure \ref{fig:Deposition_diagrams}b). This method requires the selected organic separation solvent to be more volatile than water so that upon pouring, it evaporates rapidly leaving the interfacially assembled 2DLM platelet film in contact with the substrate, with a volume of water above. Once the separation solvent fully evaporates the film adheres to the substrate, and the water is removed from above the film by syringe or by evaporation. However, unlike the vertical and horizontal dipping approaches, this method leads to trace amounts of surfactant remaining on the substrate after the separation solvent has evaporated, whereas in the dipping approaches it is retained within the separation solvent when the substrate is removed. This can be seen in Figure \ref{fig:coffee_rings} which shows a scanning electron microscope (SEM) image of a graphene film deposited by pouring on copper, where characteristic `coffee rings' showing residual surfactant can be seen, which are not present on dipped films.\\

\begin{figure}[t]
    \centering
    \includegraphics[width=\linewidth]{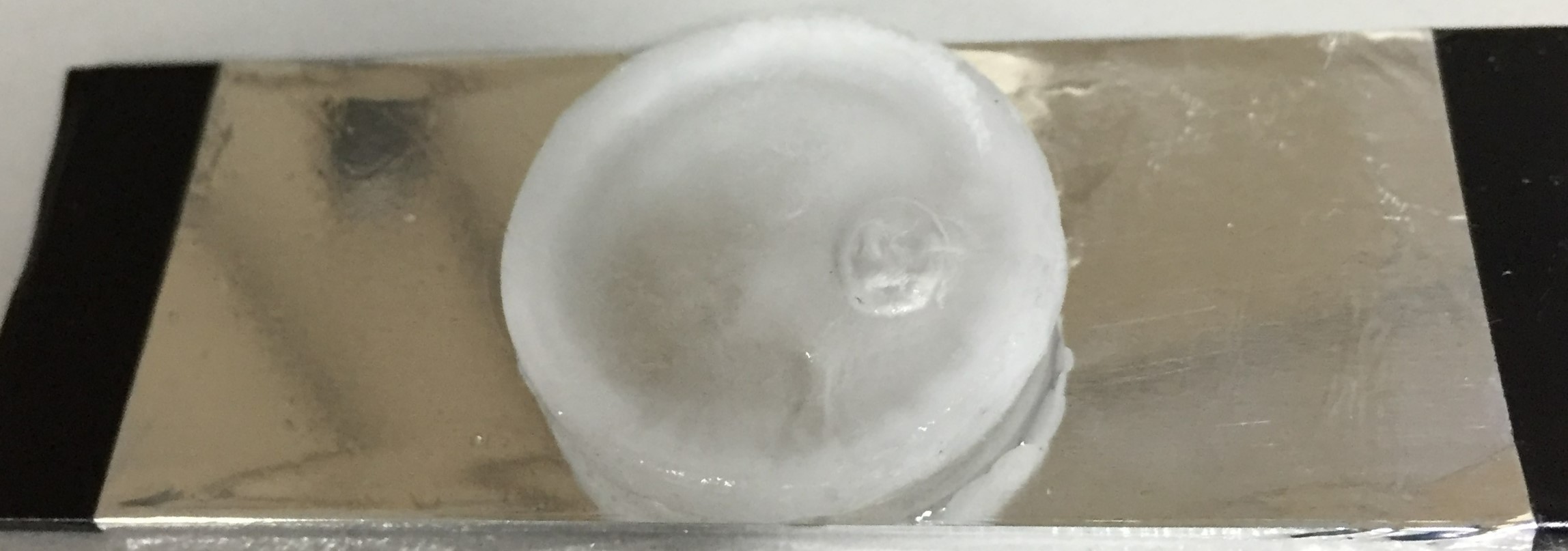}
    \caption{\textbf{Photograph of stage in freezing method.} Photograph of second stage of freezing method of deposition, after the separation solvent has evaporated bringing the 2DLM platelet film frozen within the water droplet into contact with the substrate. The ice is then allowed to melt and remaining water is removed via evaporation or using a syringe, leaving the 2DLM platelet film deposited on the substrate. The thin cylinder of ice has a diameter of 2.5~cm.}
    \label{fig:freeze_transfer}
\end{figure}

It is also possible to transfer the interfacially assembled 2DLM platelet film from the centrifuge tube to a suitable substrate by a novel approach which we term `Freeze Transfer'. In this method the centrifuge tube containing the interface assembled 2DLM platelet film is cooled to below the freezing point of both liquid phases, immobilising the interface and 2DLM film. The frozen block of material is then placed with the separation solvent in contact with the sample. The separation solvent melts more rapidly than the ice, bringing the film into contact with the substrate (Figure \ref{fig:freeze_transfer}). Once the ice has melted, the water is then removed by evaporation or by means of syringe.

\clearpage

\begin{figure}[h]
    \centering
    \includegraphics[width=0.75\columnwidth]{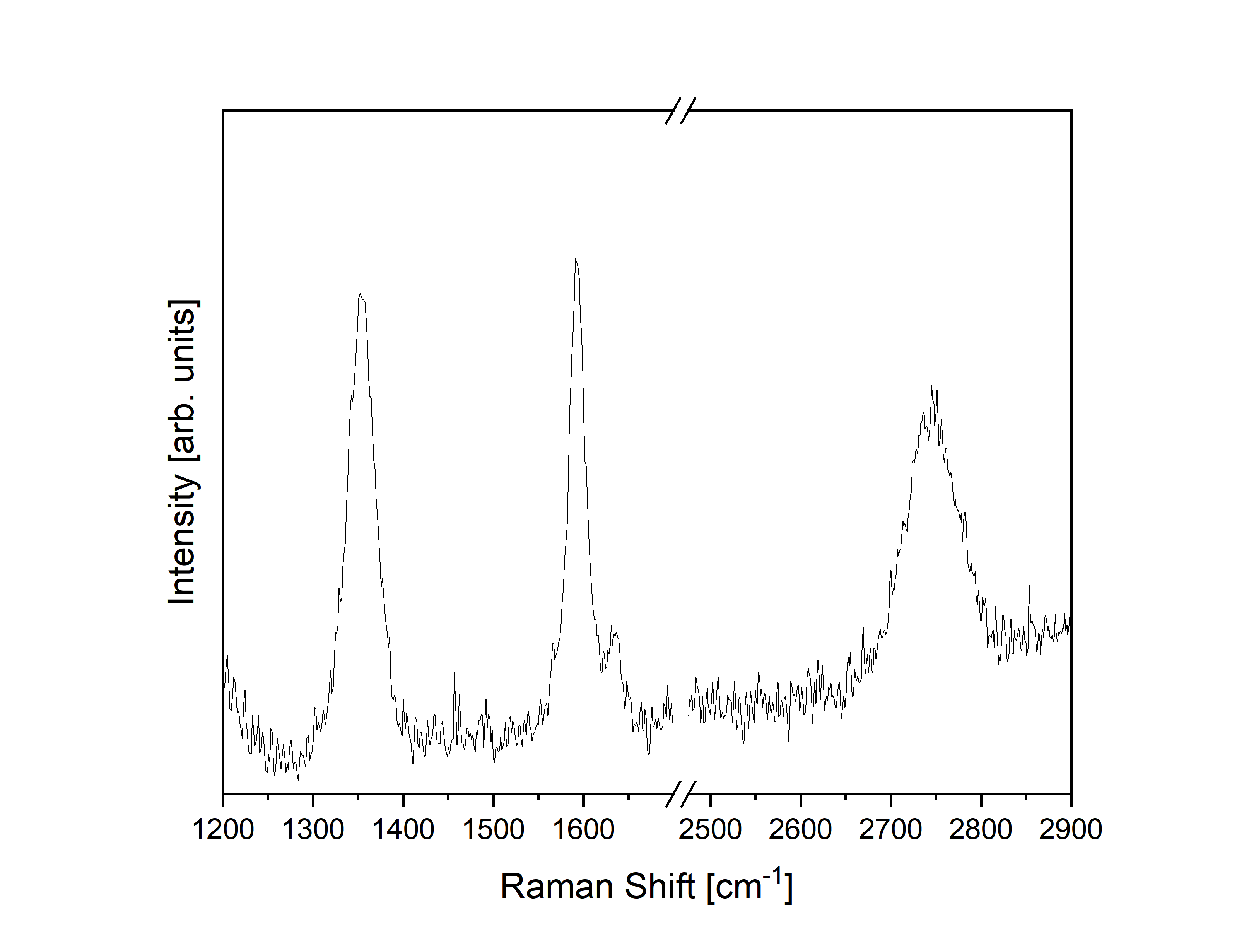}
    \caption{\textbf{Graphene film using alternative surfactant.} Raman spectrum of a few-layer graphene film $\sim$ 12 layers thick made using the LID method from a Tween 20 stabilized suspension.}
    \label{fig:tween}
\end{figure}

\begin{figure}[h]
    \centering
    \includegraphics[width=0.75\columnwidth]{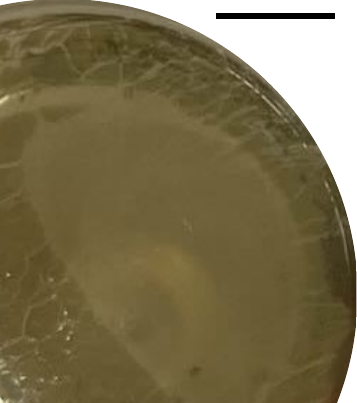}
    \caption{\textbf{MoS$_2$ film left to settle via gravity.} Photograph, viewed from above, of a MoS$_2$ film formed at the interface between water and dichloromethane after the surfactant-stabilized aqueous suspension containing MoS$_2$ platelets and dichloromethane have been mixed and allowed to settle for $\sim$27 hours. The scale bar is 5~mm.}
    \label{fig:enter-label}
\end{figure}

\clearpage

\begin{figure}[h]
    \centering
    \includegraphics[width=0.75\linewidth]{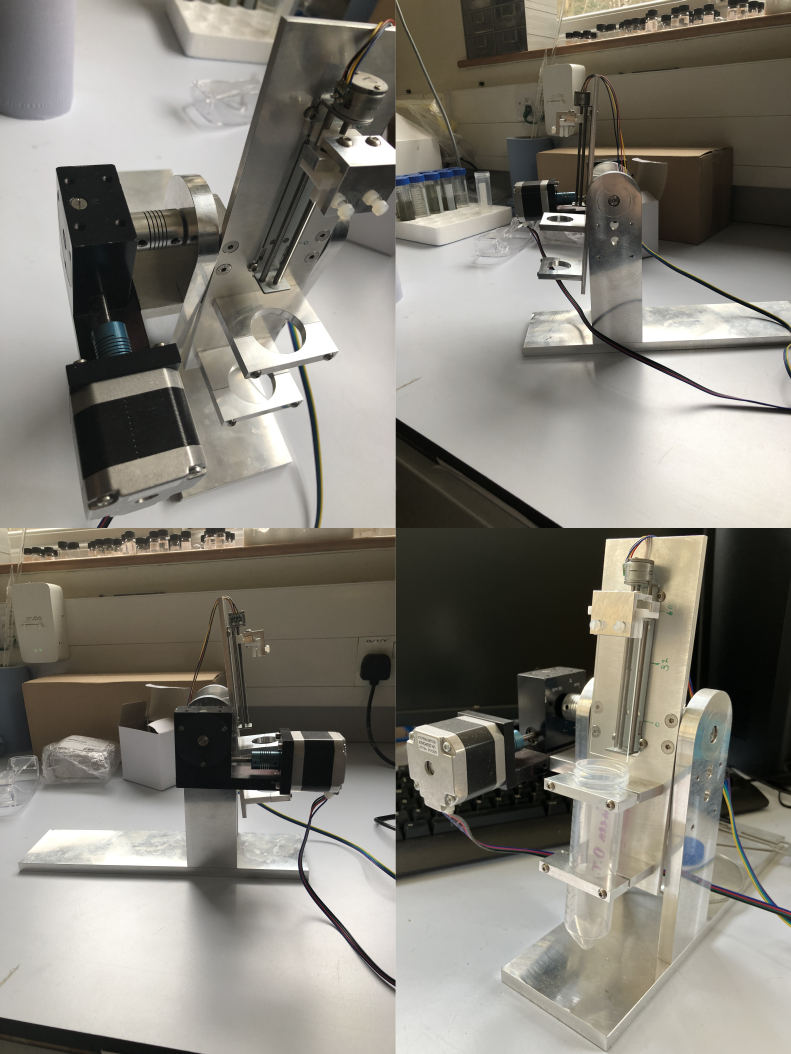}
    \caption{\textbf{Photograph of machine for automated dipping.} Photographs of machine built to automate the dipping method. The lower right image shows the machine with a centrifuge tube and glass slide, as for deposition.}
    \label{fig:enter-label}
\end{figure}

\clearpage
\section{\raggedright{Additional Thin Films}}

\begin{figure}[h]
    \centering
    \includegraphics[width=0.75\columnwidth]{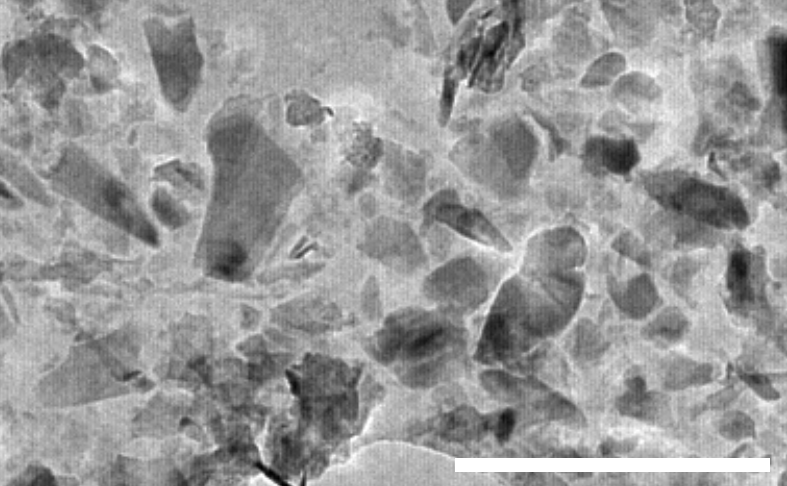}
    \caption{\textbf{MoS$_2$ transmission electron microscopy.} Transmission electron microscope image of MoS$_2$ on a holey carbon grid, where the continuity of the film can clearly be seen. The scale bar is 0.5~$\mu$m.}
    \label{fig:enter-label}
\end{figure}

\begin{figure}[h]
    \centering
    \includegraphics[width=0.75\columnwidth]{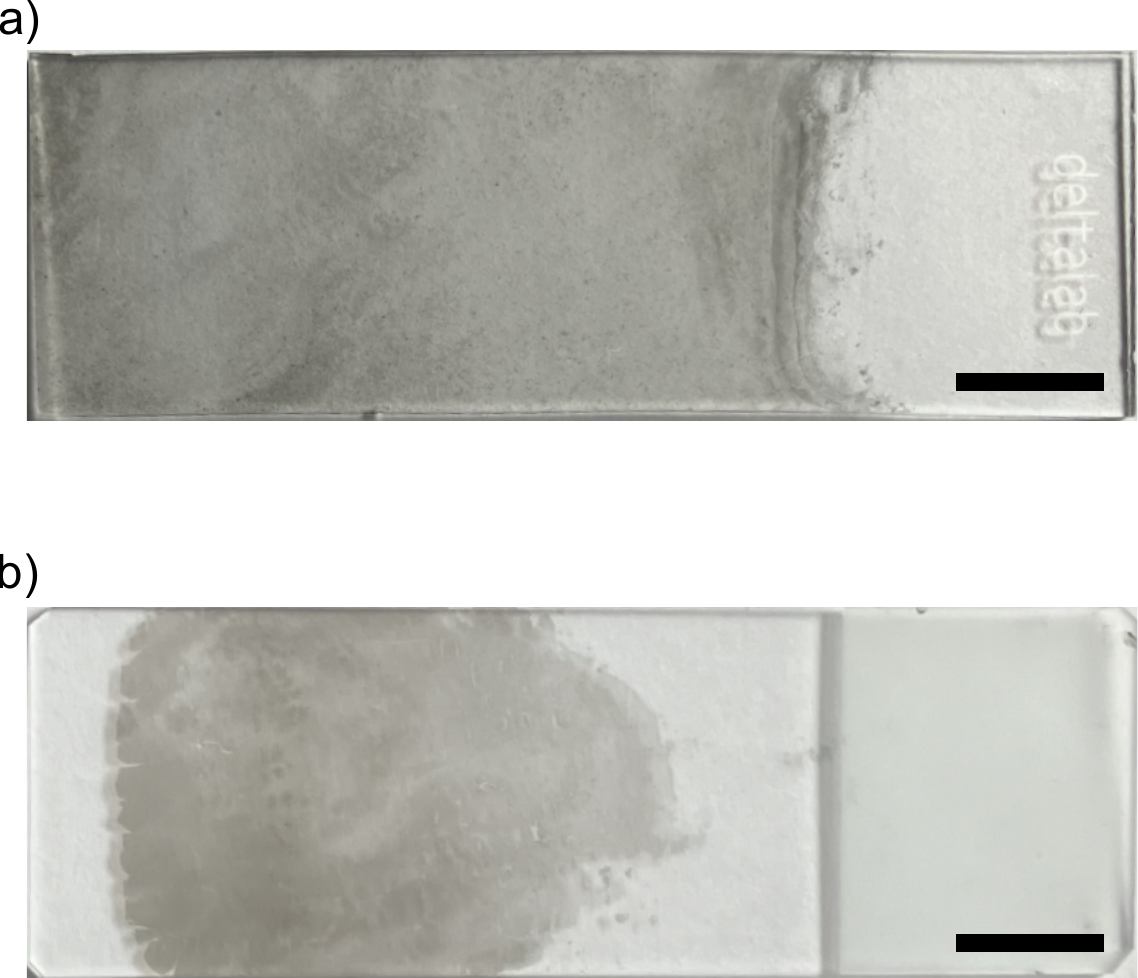}
    \caption{\textbf{Large area graphene films.} a) Large area graphene film deposited on glass by dipping multiple times. b) Large area graphene film deposited on glass by dipping once. The scale bar is 1~cm for both images.}
    \label{fig:enter-label}
\end{figure}

\begin{figure}[h]
    \centering
    \includegraphics[width=0.75\columnwidth]{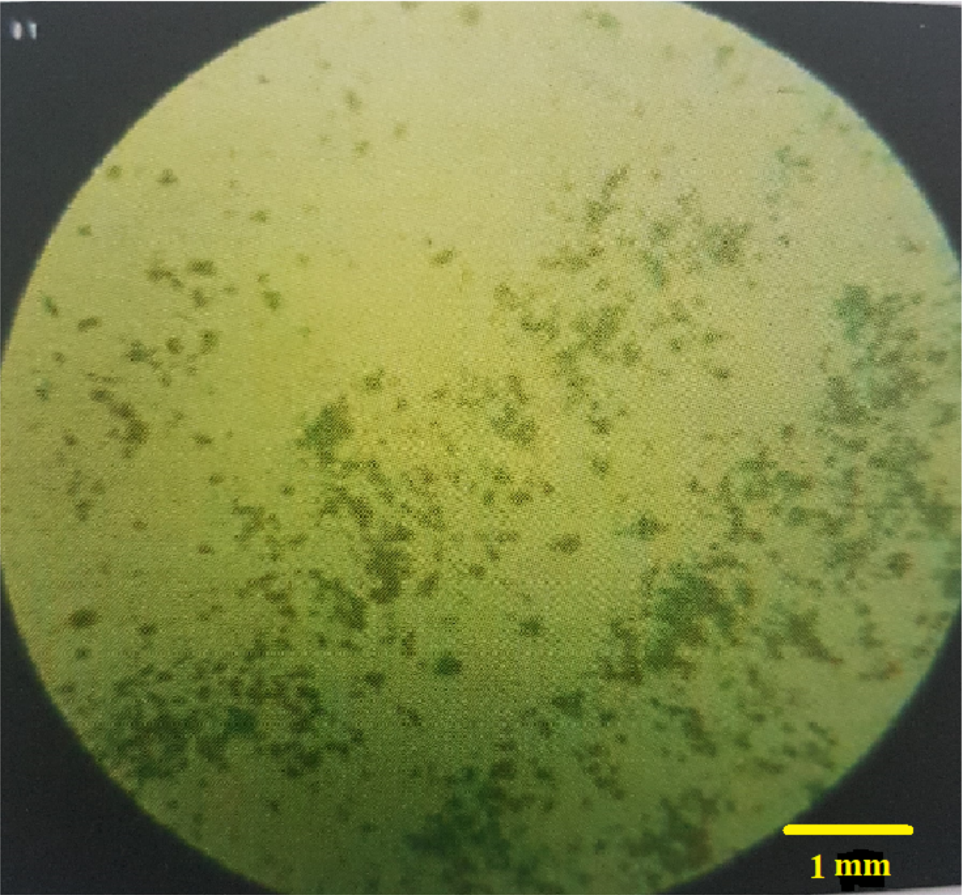}
    \caption{\textbf{Langmuir-Blodgett few-layer graphene film.} Optical image of a few-layer graphene film on glass produced using the Langmuir-Blodgett technique. Surfactant-stabilized suspensions of few-layer graphene platelets were centrifuged to attempt to remove the surfactant, and then re-suspended in dichloromethane (DCM). This suspension was deposited evenly across a water surface, the DCM allowed to evaporate, leaving the few-layer graphene platelets at the water-air interface. The graphene on the surface was compressed by means of barriers to form a film, and then deposited onto a glass slide, on the up-stroke (Z-type deposition). The film is inhomogeneous as it was produced from a surfactant-stabilized suspension, and there is no mechanism for fully removing the surfactant, unlike in the Liquid Interface Deposition method.}
    \label{fig:LB}
\end{figure}

\clearpage

\section{\raggedright{Percolation-Like Conductivity Regime in few-layer graphene films}}
\begin{figure}[h]
    \centering
    \includegraphics[width=0.75\columnwidth]{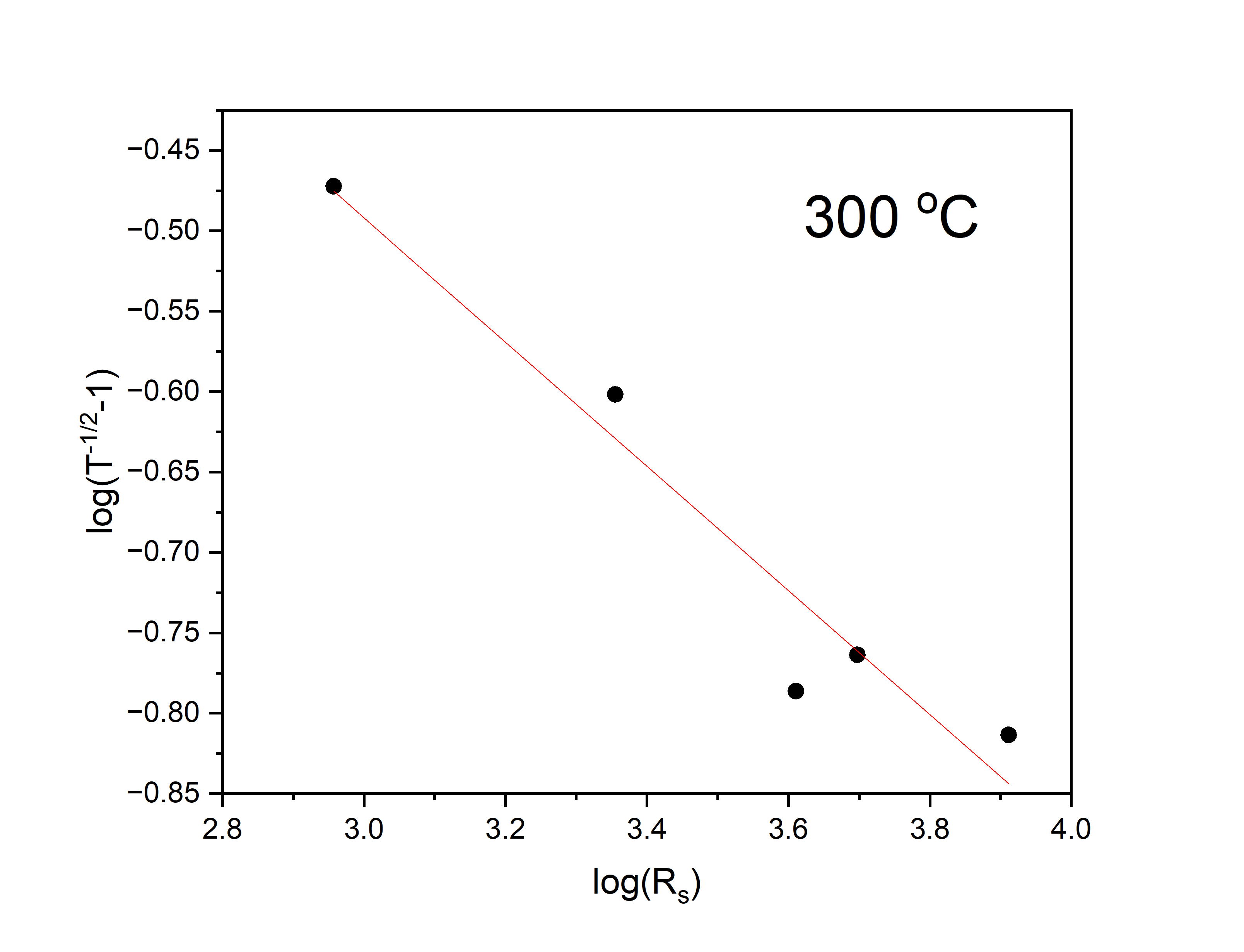}
    \caption{\textbf{Graph showing film is in percolative regime.} A plot of log($T^{-1/2}-1$) against log($R_S$) for a few-layer graphene film after the highest anneal temperature studied (300~{\textdegree}C) yields a straight line, showing this film is in the percolative-like regime.}
    \label{TvR}
\end{figure}

\begin{figure}
    \centering
    \includegraphics[width=0.75\columnwidth]{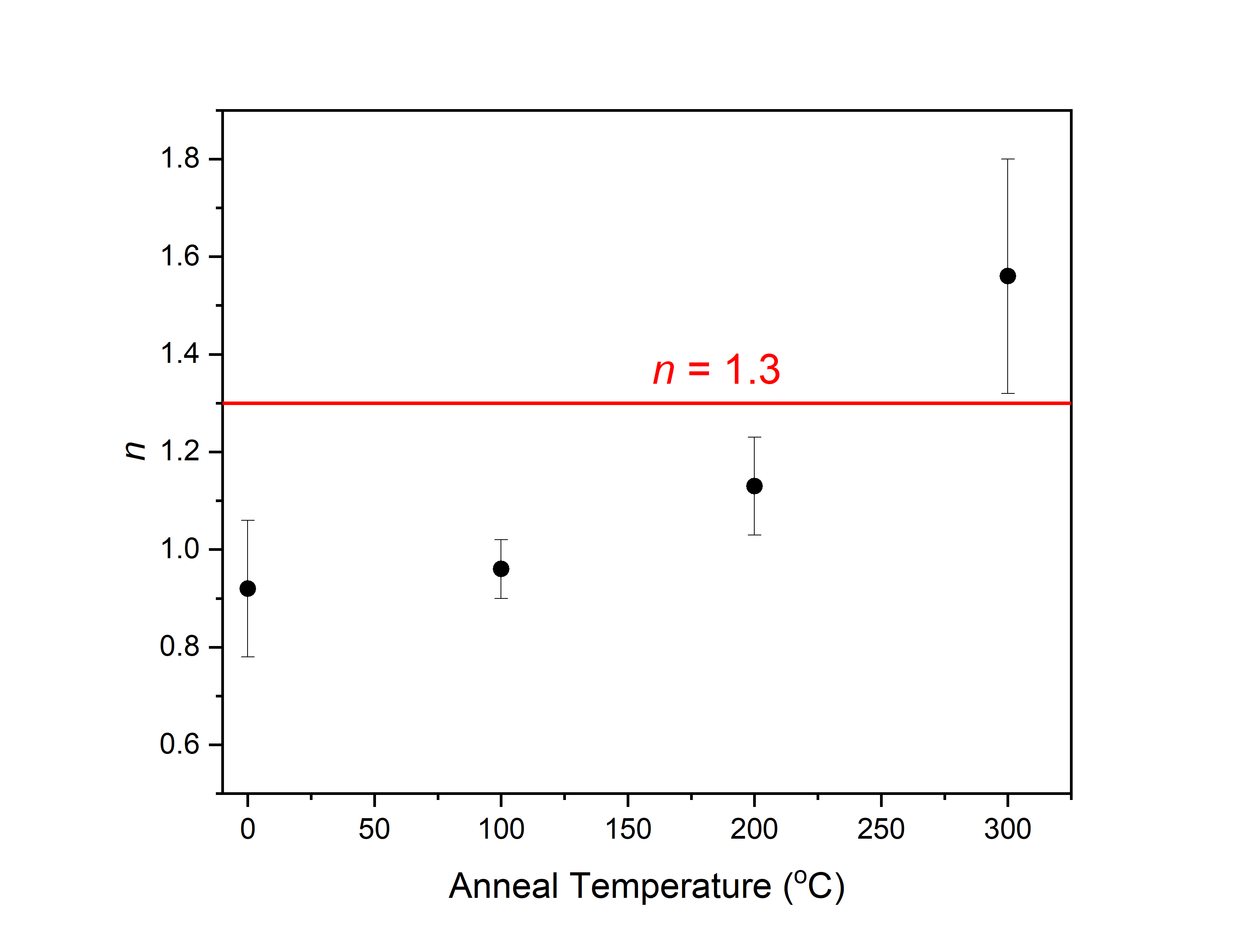}
    \caption{\textbf{Percolation exponents calculated for films after different anneal temperatures.} Non-universal percolation exponents \cite{de2010percolation} derived from transmittance and sheet resistance data from the few-layer graphene film presented in Figure 3 of the main paper as a function of anneal temperature. The universal percolation exponent ($n=1.3$) for a two-dimensional system is shown as the red horizontal line.}
    \label{fig:exp}
\end{figure}

As described in the main paper, the relationship between sheet resistance and transmission did not follow the same trend as observed in the work of De \textit{et al.}~\cite{De2010}, which would be expected for films in the bulk charge transport regime. However, subsequent work by De \textit{et al.}~\cite{de2010percolation} demonstrated a `percolation-like regime', in which films of conductive nanomaterials (including 2DLMs) have passed the conventional percolation threshold for conductivity but still display a relation between conductivity and thickness (or transmittance) which has percolation type characteristics. This regime successfully describes the behaviour of the few-layer graphene platelet films presented in the main paper, and can be described by the following equation:
\begin{equation}
    T=\bigg(1+\frac{1}{\Pi} \bigg(\frac{Z_0}{R_S}\bigg)^{1/(n+1)}\bigg)^{-2},
    \label{1}
\end{equation}
where $T$ is the transmission, $Z_0$ is the impedance of free space (377~$\Omega$), $R_S$ is the sheet resistance, $n$ is a non-universal percolation exponent, and $\Pi$ is the percolative figure of merit:
\begin{equation}
    \Pi = 2 \bigg[\frac{\sigma_{DC,B}/\sigma_{Op}}{(Z_0 t_{min} \sigma_{Op})^n}\bigg]^{1/(n+1)}.
\end{equation}
$\sigma_{DC,B}$ is the bulk DC conductivity, $\sigma_{Op}$ is the optical conductivity, and $t_{min}$ is the thickness at which the bulk conductivity is reached. Rearranging Equation \ref{1}, it can be seen that a log-log plot of $(T^{-1/2}-1)$ against $R_S$ should give a straight line if the film is in the percolation-like regime. This relationship is observed in our data for all film annealing temperatures and that for highest anneal temperature is presented in Figure \ref{TvR}. The percolation exponent, $n$ can be determined from the gradient of the straight-line fit and values are plotted for all  the annealling  temperatures explored in this study in Figure \ref{fig:exp}. The observed percolation exponents are close to the value of the universal percolation exponent for a two-dimensional system ($n=1.3)$ (shown as a red line), suggesting the films exhibit a high degree of two-dimensional behaviour.

\newpage
\clearpage
\section{\raggedright{Further Supplementary Data}}
\begin{figure}[h]
    \centering
    \includegraphics[width=\columnwidth]{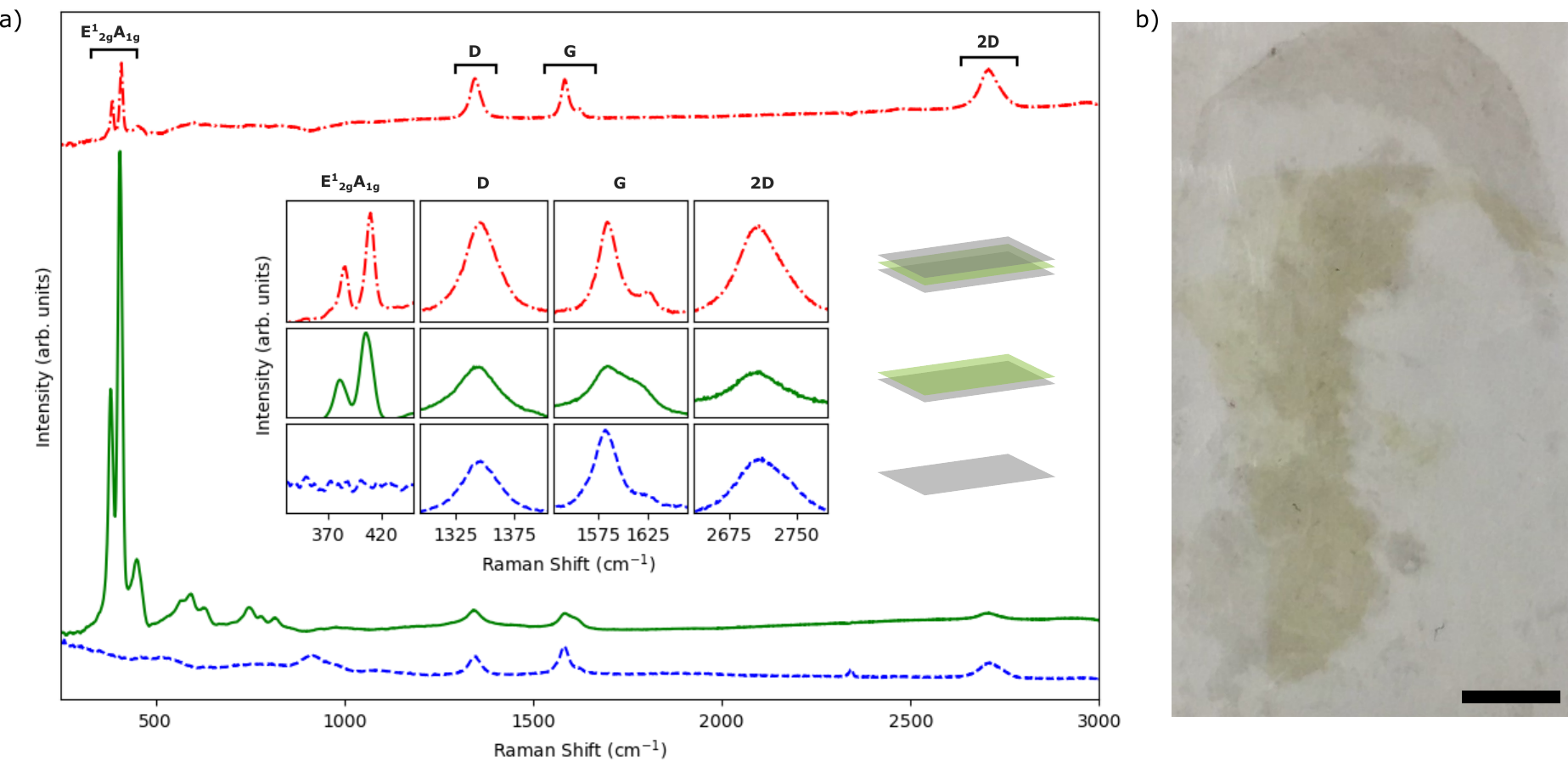}
    \caption{\textbf{Graphene-MoS$_2$-graphene van der Waals heterostructure.} a) Raman spectra of a few-layer graphene (FLG)-MoS$_2$-FLG vdWH produced via the pouring method for each fabrication step. The blue data (dashed line) correspond to pure FLG, the green data (continuous line) to a two-layer FLG-MoS$_2$ vdWH, and the red data (dash-dot line) to a trilayer FLG-MoS$_2$-FLG vdWH. The data have been offset vertically to allow for comparison, and the insets highlight regions containing the main Raman lines of the two materials. Features from both MoS$_2$ and graphene are present in the Raman spectra of the final vdWH. b) Photograph of the whole glass slide containing the sample. Areas containing the FLG-MoS$_2$-FLG vdWH, can be seen alongside the regions coated only with FLG or the two-layer vdWH. The latter are distinguished unambigiously from the trilayer by Raman spectroscopy. The scale bar is 5~mm.}
    \label{fig:heterostructure}
\end{figure}

\begin{figure}[h]
    \centering
    \includegraphics[width=0.75\columnwidth]{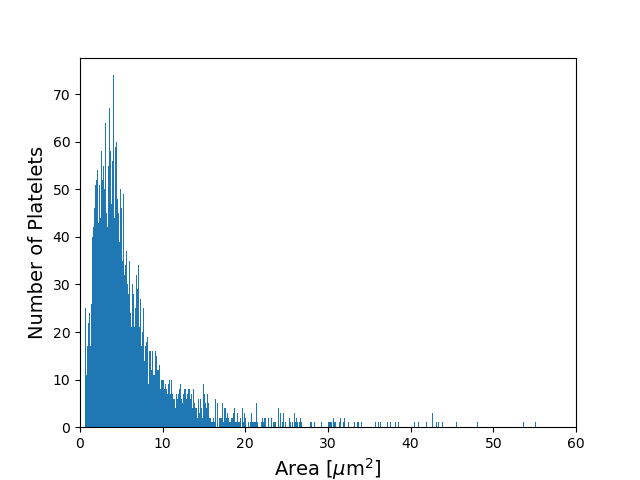}
    \caption{\textbf{Platelet area distribution for a MoS$_2$ suspension.} Area distribution for a typical sample of surfactant-stabilized MoS$_2$ platelets used to produce films presented in this work. Areas were calculated from Scanning Electron Microscope images of  MoS$_2$ platelets vacuum filtered onto a cellulose nitrate membrane from suspension. Approximately 3800 platelets were used to produce the distribution shown.}
    \label{fig:MoS2 analysis}
\end{figure}

\newpage
\clearpage

\begin{sidewaystable}
\centering
    \begin{tabular}{| p{4.6cm} | p{4.3cm} | p{6.3cm} | p{2.1cm} | p{2cm} | p{3.3cm} |}
        \hline
        & & & \multicolumn{2}{c|}{\textbf{Generalisable?}} & \\
        \hline
         \textbf{Reference} & \textbf{Fabrication method} & \textbf{Film deposition method} & \textbf{Fabrication} & \textbf{Deposition} & \textbf{`Green' technique?}\\
         \hline
         This work & High-shear exfoliation & Liquid Interface Deposition & Yes & Yes & Yes\\
         \hline
         Mojtabavi \textit{et al.}~\cite{mojtabavi2021wafer} & Salt-acid etching & Self-assembly at interface & No & No & No\\
         \hline
         Shi \textit{et al.}~\cite{shi2024ultrafast} & Various & Interfacial assembly assisted by surfactant & No & Yes & No\\
         \hline
         Li \textit{et al.}~\cite{li2014inkjet} & Sonication & Inkjet printing & Yes & Yes & No\\
         \hline
         Kim \textit{et al.}~\cite{kim2021solution} & Tip sonication & Vacuum filtration followed by annealing & Yes & No & No\\
         \hline
         Kim \textit{et al.}~\cite{kim2022all} & Electrochemical exfoliation & Spin coating & No & Yes & No\\
         \hline
         Cunningham \textit{et al.}~\cite{cunningham2013photoconductivity} & Tip sonication & Lagmuir-Blodgett based method & Yes & Unclear & Yes\\
         \hline
         Abdolhoss-einzadeh \textit{et al.}~\cite{abdolhosseinzadeh2022universal} & Bath sonication & Multiple printing methods & Yes & Yes & No\\
         \hline
         Higgins \textit{et al.}~\cite{higgins2019electrolyte} & Probe sonicated & Airbrush spraying & Yes & Unclear & Yes\\
         \hline
         Zou \textit{et al.}~\cite{zou2023high} & Electrochemical exfoliation & Drop casting & No & Yes & No\\
         \hline
    \end{tabular}
\captionof{table}{\textbf{Comparison between different methods of producing thin films of 2D material.}}
\label{table:S1}
\end{sidewaystable}

\newpage
\clearpage

\bibliography{SuppMat}